\documentclass[sn-mathphys,Numbered]{sn-jnl}% Math and Physical Sciences Reference Style
\usepackage{graphicx}%
\usepackage{multirow}%
\usepackage{amsmath,amssymb,amsfonts}%
\usepackage{amsthm}%
\usepackage{mathrsfs}%
\usepackage[title]{appendix}%
\usepackage{xcolor}%
\usepackage{textcomp}%
\usepackage{manyfoot}%
\usepackage{booktabs}%
\usepackage{algorithm}%
\usepackage{algorithmicx}%
\usepackage{algpseudocode}%
\usepackage{listings}%
\theoremstyle{thmstyleone}%
\theoremstyle{thmstyletwo}%

\theoremstyle{thmstylethree}%

\def\be{\begin{equation}}
\def\ee{\end{equation}}
\def\ba{\begin{eqnarray}}
\def\ea{\end{eqnarray}}
\def\nn {\nonumber}
\def\bea {\begin{eqnarray}}
\def\eea {\end{eqnarray}}
\def\be {\begin{equation}}
\def\ee {\end{equation}}
\def\nn {\nonumber}
\def\red {\color{red}}

\begin{document}

\title[Empirical determination of the energy loss of heavy quarks in nuclear collisions at RHIC and LHC energies]{Empirical determination of the energy loss of heavy quarks in nuclear collisions at RHIC and LHC energies}


\author[1]{\fnm{Somnath} \sur{De}}\email{somvecc@gmail.com}

\author*[2]{\fnm{Sudipan} \sur{De}}\email{sudipan86@gmail.com}

\author[3]{\fnm{Prashant} \sur{Shukla}}\email{pshuklabarc@gmail.com}

\affil[1]{\orgdiv{Department of Physics}, \orgname{Pingla Thana Mahavidyalaya}, \orgaddress{\street{Malligram}, \city{Paschim Medinipur}, \postcode{721140}, \state{West Bengal}, \country{India}}}

\affil*[2]{\orgdiv{Department of Physics}, \orgname{Dinabandhu Mahavidyalaya}, \orgaddress{\street{Bongaon}, \city{North 24 Parganas}, \postcode{743235}, \state{West Bengal}, \country{India}}}

\affil[3]{\orgdiv{Nuclear Physics Division}, \orgname{Bhabha Atomic Research Center}, \orgaddress{\street{Trombay}, \city{Mumbai}, \postcode{400085}, \state{Maharashtra}, \country{India}}}

\affil[3]{\orgname{Homi Bhabha National Institute}, \orgaddress{\street{Anushakti Nagar}, \city{Mumbai}, \postcode{400094}, \state{Maharashtra}, \country{India}}}


\abstract{Heavy quarks produced in the heavy ion collisions loose energy while propagting in the
hot partonic matter which finally fragment to heavy ($D$ or $B$) mesons.
The energy loss suffered by the heavy quarks is imprinted in the nuclear modification factor
as a function of transverse momenta ($p_T$) of these heavy mesons.
 An alternate measure of in-medium energy loss comes through the effective shift in
transverse momentum spectra of hadrons recorded in nucleus-nucleus collisions when
it is compared to the same in proton-proton collisions.
We start by parametrizing invariant momentum yields of heavy mesons in p+p collisions.
The fit function from p+p collisions and measured nuclear modification factor in
heavy ion collisions are then utilised to obtain the shift in the transeverse
mass $\Delta m_T$ of heavy mesons produced at the RHIC and LHC experiments.
The energy loss $\Delta m_T$ so obtained is found to scale with the transverse mass ($m_T$) of
heavy mesons through a power law at different energies and centralities of collisions.
We have also calculated using theoretical formalism, the total energy loss suffered
by a charm quark in quark-gluon plasma produced in Pb~+~Pb collisions at the LHC
energies. The evolution of the plasma is described by (2+1) dimensional longitudinal
boost-invariant ideal hydrodynamics. It is found that the total energy loss of charm
quarks scales with the transverse mass of charm quarks through a similar power law
which supports our empirical analysis of energy loss.}




\maketitle

\section{Introduction}\label{sec1}
\label{intro}

Ultra-relativistic heavy ion collision experiments at BNL and CERN facilities have
succeeded to create a strongly interacting, hot and dense, deconfined state of nuclear
matter of energy density $\sim$ 0.4 GeV/$fm^3$ and of temperature about few hundreds of MeV~\cite{QGP-review_23}.
The wealth of data from RHIC (Relativistic Heavy Ion Collider)~\cite{Rhic-white}
and LHC (Large Hadron Collider)~\cite{QM-12,Rev, Harris}
experiments during past few decades provide strong evidences for the existence
of such matter, which is known as Quark-Gluon Plasma (QGP).
At length scales much larger than the characteristic mean free path of partons,
QGP behaves like a strongly correlated liquid and is best described by relativistic
hydrodynamics (ideal or viscous). However at length scales much smaller than the mean
free path, QGP is considered as weakly interacting gas of partons and 
the interactions among themselves is governed by perturbative Quantum ChromoDynamics
(pQCD)~\cite{Jeon_hydro}.
 Several signatures of QGP, \textit{viz.}, elliptic and higher order anisotropic flow,
jet-quenching, radiation of photons and dileptons, suppression and regeneration
of heavy quarkonia; sensitive to both short and long distance dynamics 
have been confirmed~\cite{signature_qgp, Review_18,Quarkonia}. 
  In the present work, we shall discuss about high transverse momentum ($p_T$)
probes of QGP namely, \textit{heavy quarks}. Heavy quarks are especially useul because they are produced
in the earliest stage of the collision and their momentum spectra could reveal
the mechanism of energy loss in the QGP medium.
While light quark and gluon jets predominantly lose energy through gluon
bremsstrahlung~\cite{Baier}, heavy quark jets lose energy through elastic
collisions~\cite{BT-coll,PP-coll} as well as induced 
gluon radiation~\cite{DGLV, HT, Abir_Jamil,Kapil_NPA} in the medium.
The amount of energy loss is related to the energy density and path-length
traversed in the medium~\cite{T.Renk}. 
This phenomenon is often quantified through the nuclear modification factor $(R_{\textrm{AA}}(p_T))$ of produced 
hadrons from jets~\cite{X.N.Wang,jet-quench}.
It is defined as the ratio of invariant momentum yield of hadrons in nucleus-nucleus (AA)
collisions and the same in elementary proton-proton (pp) collisions, multiplied
by nuclear overlap function ($T_{\textrm{AA}}(b)$). The observation of $R_{AA}$ less
than unity, strongly indicates the creation of a interacting partonic medium apart from 
the contribution of cold nuclear matter effect~\cite{CNM}.
This can be understood qualitatively as follows: high $p_T$ partons produced after
initial scattering, have steeply falling momentum spectra.
Now they lose energy via multiple collisions and radiation while traversing the medium
which results in deficit of parton yield in a given $p_T$ bin. The effect 
will be revealed on the momenta of hadrons, fragmenting from the parton.
Several studies have been carried out in order to shed light on the system size
and collision energy dependence of 
parton energy loss~\cite{GLV,XNWang2,Yajem,S_dk1,S_dk2,Younus_dk}. 
However it came as a surprise when the measured $R_{AA}$ of light hadron
production in Au~+~Au collisions at RHIC~\cite{RHIC-pi} and Pb~+~Pb collisions at
the LHC~\cite{ALICE-charge} energies are found quite similar. 
This fact motivated us to calculate \textit{energy loss} which had been advocated 
earlier in refs.~\cite{delpt_wang,Horowitz, Phenix-eloss}.
The empirical study reported in ref.~\cite{delpt_wang} has obtained the fractional
energy loss ($\Delta p_T/ p_T$) of high $p_T$ charged particles, neutral pions
and non-photonic electrons for Au~+~Au collisions at the RHIC energy. 
The authors of ref.~\cite{Horowitz} have shown that the fractional energy loss
of quarks and gluons for a longitudinally expanding plasma are comparable at RHIC and 
LHC kinematic conditions.
Later, the PHENIX Collaboration~\cite{Phenix-eloss} have reported the scaling
properties of fractional energy loss ($\Delta p_T/ p_T$) of light hadrons from
the invariant yield measurements in p+p and A+A collisions at the center of mass
energies ($\sqrt{s_{NN}}$) varying from 62.4 GeV to 2.76 TeV. There are also recent
studies which have employed a similar apporach to the inclusive charged particle or
inclusive jet measurement at the LHC energies~\cite{Antonio_prc, Martin_epj, Kapil_JPG}.

In this work, we extract the energy loss of heavy mesons using measurements of
their $R_{\textrm{AA}}$ in nuclear collisions at
energies ranging from $\sqrt{s_{NN}}=$ 200 GeV to  5.02 TeV.
For this purpose, we have parameterised heavy meson ($D^0$ and $B^+$) $p_T$ spectra in
p+p collisions with Hagedorn distribution~\cite{Hagedorn1}. Then we obtain the
energy loss parameter $\Delta m_T$ in terms of measured $R_{\textrm{AA}}$
and the parameters of Hagedorn function.
The paper is organised as the follows. In the next section, we have described the methodology
adopted in this work. In section 3, we give the Hagedorn fit function of the $p_T$
spectra in p+p collisions and obtained the behaviour of $\Delta m_T$ as a function
of $m_T$ for heavy ion collisions.
In section 4, we employed theoretical model to calculate the energy
loss ($\Delta E_T$) of a heavy quark for a given path length inside QGP and obtained
$\Delta E_T$ as a function of $E_T$ for different formalism of radiative
energy losses. Finally, we have summarised the study in section V.

\section{Formalism}
%%%%%%%%%%%%%%%%%%%%%%%%%%%%%%%%%%%%%%%%%%%%%%%%%%%%
The nuclear modification factor ($R_{\textrm{AA}}$) of hadron production is
often described as the ratio of production cross section of hadrons in A+A collisions,
scaled to each nucleon-nucleon collision and the production cross-section of hadrons in p+p collisions,
%%%%%%%%%%%%%%%%%%%%%%%%
\be
R_\textrm{AA} (p_T, b) = \frac{d\sigma^{\textrm{AA}}/ d^2p_T dy (b)}{d\sigma^{\textrm{pp}}/ d^2p_T dy}.
\label{Raa}
\ee
%%%%%%%%%%%%%%%%%%%
In the 80's, Hagedorn proposed an empirical function ~\cite{Hagedorn1} which successfully describes
the momentum distribution of particles produced in high energy hadron-hadron collisions:
%%%%%%%%%%%%%%%%%%%%%%%%%
\be
d\sigma / d^2p_T dy = A\left(1+\frac{m_T}{p_0}\right)^{-n}.
\label{Hag}
\ee
%%%%%%%%%%%%%%%%%%%%%%
where $A$, $n$, $p_0$ are parameters to be fitted and $m_T= \sqrt{p_T^2+m^2}$.
We have chosen $m_T$ instead of $p_T$ as it is the relevant hard momentum scale for heavy
quark production~\cite{FONLL}.
The success of Eq.~\ref{Hag} lies in the fact that the distribution
encodes both exponential (low $p_T$ limit) and power law (high $p_T$ limit) behaviour of the particle spectra.
%%%%%%%%%%%%%%%%%%%%%%%%%%%%%%%%%%%%%
The power $n$ appears in the Eq.~\ref{Hag}, according to \textquoteleft QCD-inspired\textquoteright~quark exchange model, 
can be related to the nature of parton scatterings involved in the collisions~\cite{QCD-power, QCD-power1, QCD-power2}. 
The model suggests that the QCD cross sections scale as $1/p_T^n$, the power $n  = (2n_a - 4),$ where $n_a$  is the number of active flavors. 
When the dominant sub-process in hadron production is point like  quark-quark scatterings (referred as leading twist) 
the number of participating quarks is 4 and hence n  = 4. The power can go higher when multiple quark-quark scatterings or 
quark-hadron scatterings (referred as higher twist) are the dominant sub-processes. 
The analysis of power n is extensively studied in~\cite{P.khandai} for light hadron production in Proton-Proton collisions at 
RHIC and LHC center of mass energies.
The power $n$ is also found sensitive to the center of mass energy and the hadron species~\cite{Arleo_PRL}.
%%%%%%%%%%%%%%%%%%%%%%

%%%%%%%%%%%%%%%%%%%%%%%%%%%%%%%%%%%%%%
\begin{table*}[!]
\setlength{\arrayrulewidth}{0.20mm}
\setlength{\tabcolsep}{10pt}
\renewcommand{\arraystretch}{1.1}
%\settowidth{\columnwidth}
%\begin{widetext}
\begin{tabular}{ccccccc} \hline
\multicolumn{1}{c}{$\sqrt{s_{NN}}$} & 
\multicolumn{1}{c}{Experiment} &
\multicolumn{1}{c}{Meson} &
\multicolumn{1}{c}{A} &
\multicolumn{1}{c}{$n$} & 
\multicolumn{1}{c}{$p_0$} &
\multicolumn{1}{c}{$\chi^2/NDF$} \\
(TeV) &  &  & ($ (\mu \textup{b}) GeV^{-2}$) & & (GeV)& \\
\hline\hline
0.200 & STAR & $D^0$ & (3.00$\pm$2.82)$\times10^4$&11.77$\pm$0.70 & 2.08$\pm$0.43 & 0.042 \\
2.76 & CMS & $D^0$ &(2.15$\pm$1.77)$\times10^4$ &6.96$\pm$0.69 & 1.27$\pm$0.84 & 0.003 \\
5.02 & CMS & $D^0$ &(4.00$\pm$2.47)$\times10^4$ &6.32$\pm$0.32 & 1.03$\pm$0.17 & 0.004 \\
2.76 & ALICE  & $D^0$ &(8.53$\pm$4.87)$\times10^3$ & 6.75$\pm$0.61 & 1.04$\pm$0.34 & 0.017 \\
5.02 & ALICE & $D^0$ &(4.99$\pm$2.55)$\times10^4$& 6.13$\pm$0.43& 0.92$\pm$0.16 & 0.018 \\
5.02 & CMS & $B^+$ &(2.40$\pm$2.63)$\times10^4$& 5.79$\pm$0.75& 0.86$\pm$0.17 & 0.001 \\
\hline
%\hline
\end{tabular}
\caption{The parameters of Hagedorn function (Eq.~\ref{Hag}) when fitted to $D^0$
invariant yield in proton-proton collisions at different collision energies.
The chi-square per degrees of freedom are listed for each case.}
\label{Table_spectra}
%\end{widetext}
\end{table*}
%%%%%%%%%%%%%%%%%%%%%%%%%%%%%%%%%%%%%%%%%%%%%%%%%%%%%
Next we write the scaled invariant production cross-section, scaled to per nucleon-nucleon collisions, of hadrons in A+A collisions by shifting
$m_T$ by an amount $\Delta m_T$ as follows:
%%%%%%%%%%%%%%%%%%%%%%%%%%%%%%%%%%%%%%%%%%%%
\be
d\sigma^{\textrm{AA}}/ d^2p_T dy = A\left(1+\frac{m_T+\Delta m_T}{p_0}\right)^{-n}.
\label{AA_TS}
\ee
%%%%%%%%%%%%%%%%%%%%%%%%%%%%%%%%%%%%%%%%%
As the invariant yield of hadrons in A+A collisions is defined per nucleon-nucleon collisions, we have assumed the same normalization parameter A for both p+p and A+A collisions.
The reason behind writing Eq.~\ref{AA_TS} lies in the assumption that particle yield
at a given $m_T$ in A+A collisions would be similar to the yield of particles in p+p
collisions at $m_T+\Delta m_T$. The shift ($\Delta m_T$) 
includes the medium effect, chiefly energy loss of parent quark inside the plasma.
Our formalism quite resembles with Ref.~\cite{Kapil_JPG} which describes the light
parton energy loss at the LHC energies. The Eq.~\ref{AA_TS} can be considered
as a special case of the formula used in Ref.~\cite{Kapil_JPG}. Here we have assumed
that $\Delta p_T$ is slowly varying function of $p_T$, which is a good approximation
for heavy quarks at high $p_T$. Then Eq.~\ref{Raa}, for a given centrality of collision, can be expressed as:
%%%%%%%%%%%%%%%%%%%%%%%%%%%%%%%%%%
\bea
R_\textrm{AA} (p_T) &=& \frac{A\left(1+\frac{m_T+\Delta m_T}{p_0}\right)^{-n}}{A\left(1+\frac{m_T}{p_0}\right)^{-n}}.
\eea
%%%%%%%%%%%%%%%%%%%%%%%%%%%%%%%%%%%%%%
From the above expression, the shift $\Delta m_T$ can be obtained as: 
%%%%%%%%%%%%%
\be
\Delta m_T = \left((R_{AA}(p_T))^{-1/n}-1 \right)(p_0+m_T).
\label{delmt}
\ee
%%%%%%%%%%%%%%%%
and the corresponding shift in $p_T$ as:
\be
\Delta p_T = \frac{m_T}{p_T}\Delta m_T.
\label{delpt}
\ee
%%%%%%%%%%%%%%%%%%%%%%%%%%%%%%%%%
The corresponding error in $\Delta m_T$ is given by:
\be
\sigma_{\Delta m_T} = \Delta m_T\left(\sqrt{(\sigma_{f_1}/f_1)^2+(\sigma_{f_2}/f_2)^2}\right),
\label{err-delmt}
\ee
%%%%%%%%%%%%%%%%%%%%%%%%%%%%%%%
where, $f_1= \left((R_{AA}(p_T))^{-1/n}-1 \right)$,
$\sigma_{f_1} = (\sigma_{R_{AA}}/R_{AA})\times (1/n)\times (R_{AA})^{-1/n}$
and $f_2 = (p_0+m_T)$, $\sigma_{f_2} = \sigma_{m_T} = \sigma_{p_T}$.
The $\sigma_{R_{AA}}$ is the uncertainty in $R_{AA}$, is taken as the quadrature sum of statistical
and systematic uncertainties associated with the experimental data.
One should note that the $\Delta m_T$ is related to energy loss of parton in the QGP,
if one assumes the fragmentation effects on momentum spectra similar in p+p and A+A collisions. 
%%%%%%%%%%%%%%%%%%%%%%%%%%%%%%%%%%%%%%%%%%%%%%%%%%%%

%%%%%%%%%%%%%%%%%%%%%%%%%%%%%%%%%%%%%%%%%%%%%%%%%%%%
\begin{figure}[h]%
\centering
%\includegraphics[width=0.9\textwidth]{fig.eps}
%\begin{figure*}[t]
%\begin{center}
\includegraphics[width= 0.3\textwidth]{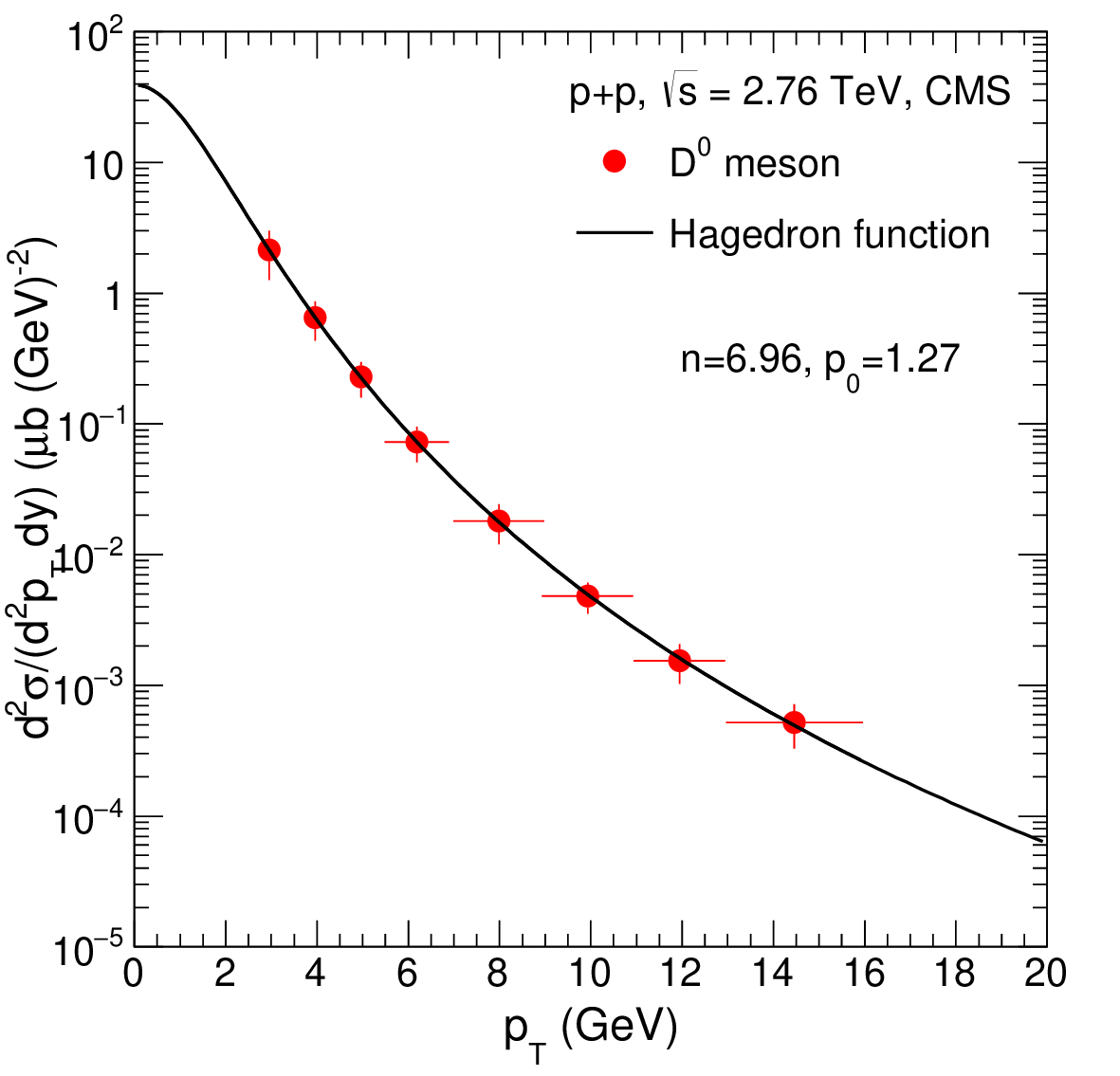}
\includegraphics[width= 0.3\textwidth]{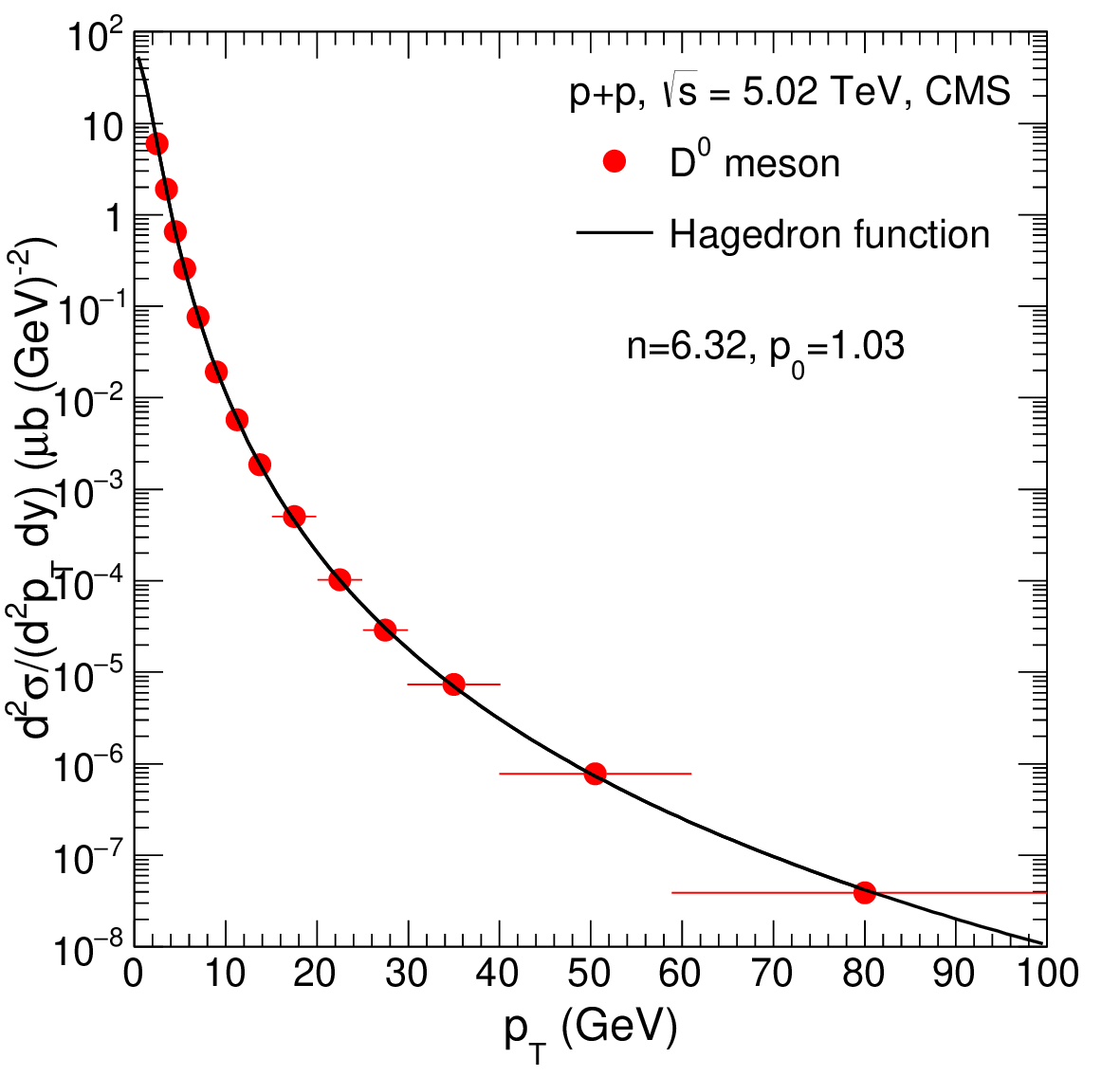}
\includegraphics[width= 0.3\textwidth]{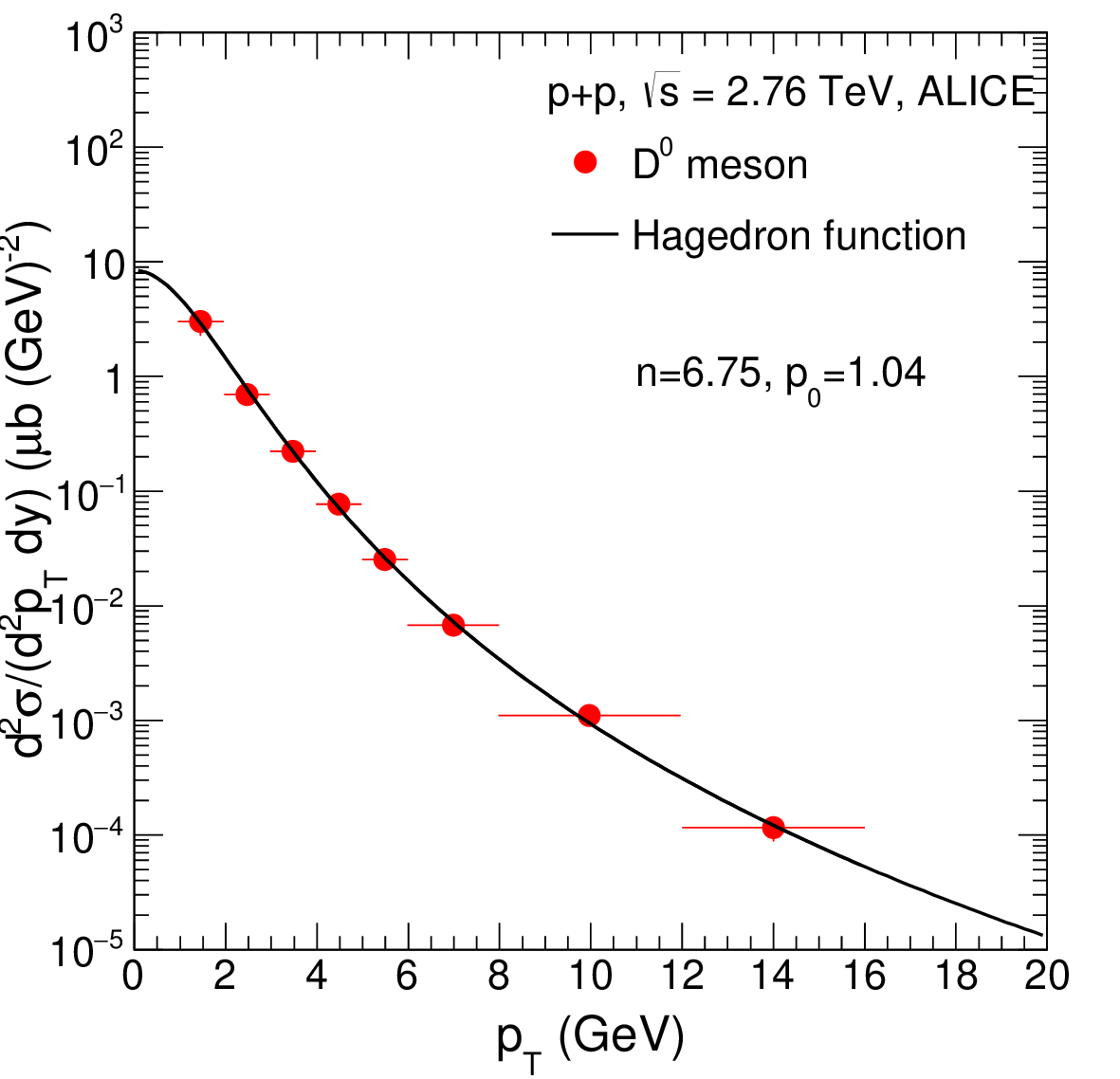}
\includegraphics[width= 0.3\textwidth]{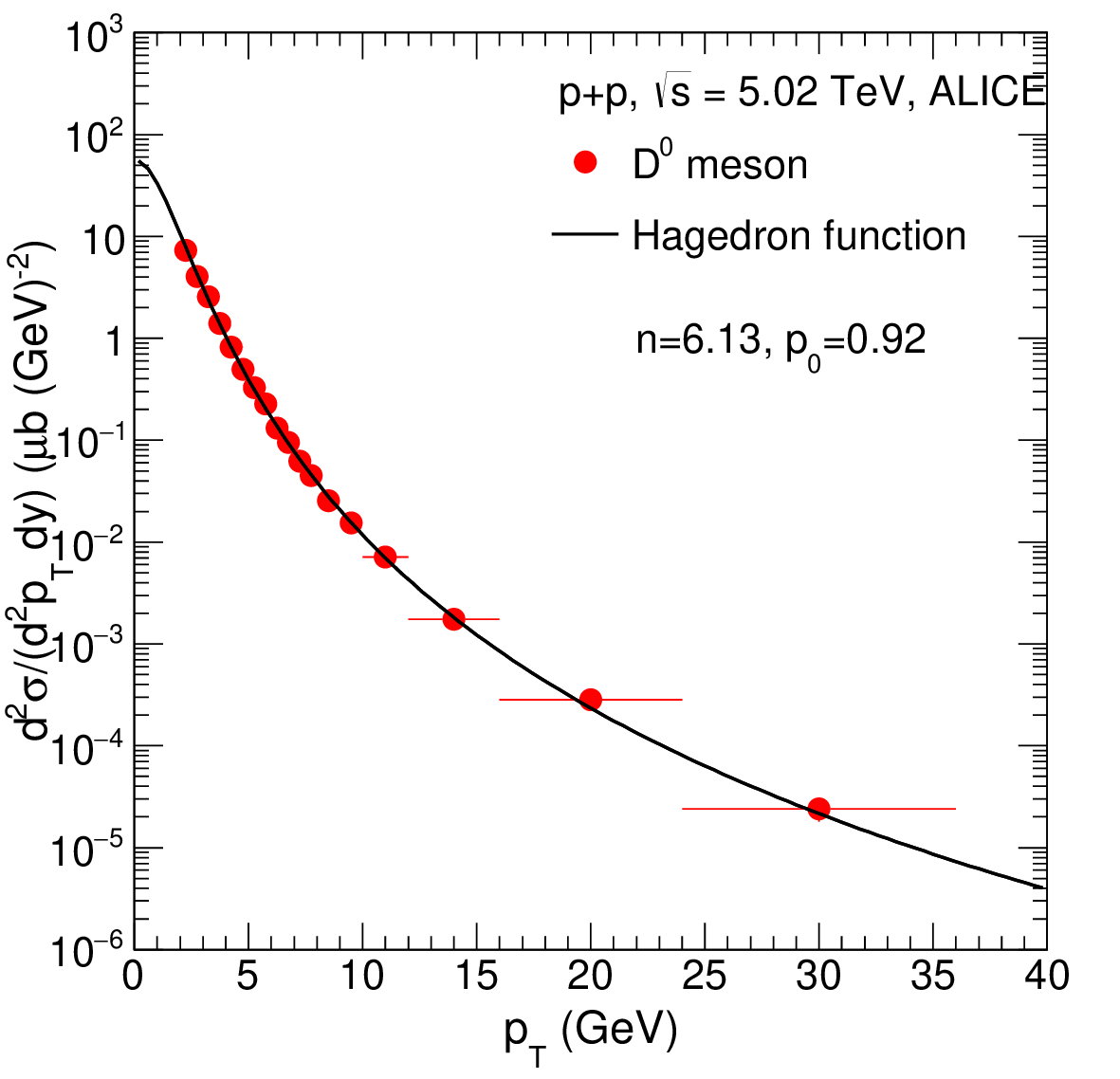}
\includegraphics[width= 0.3\textwidth]{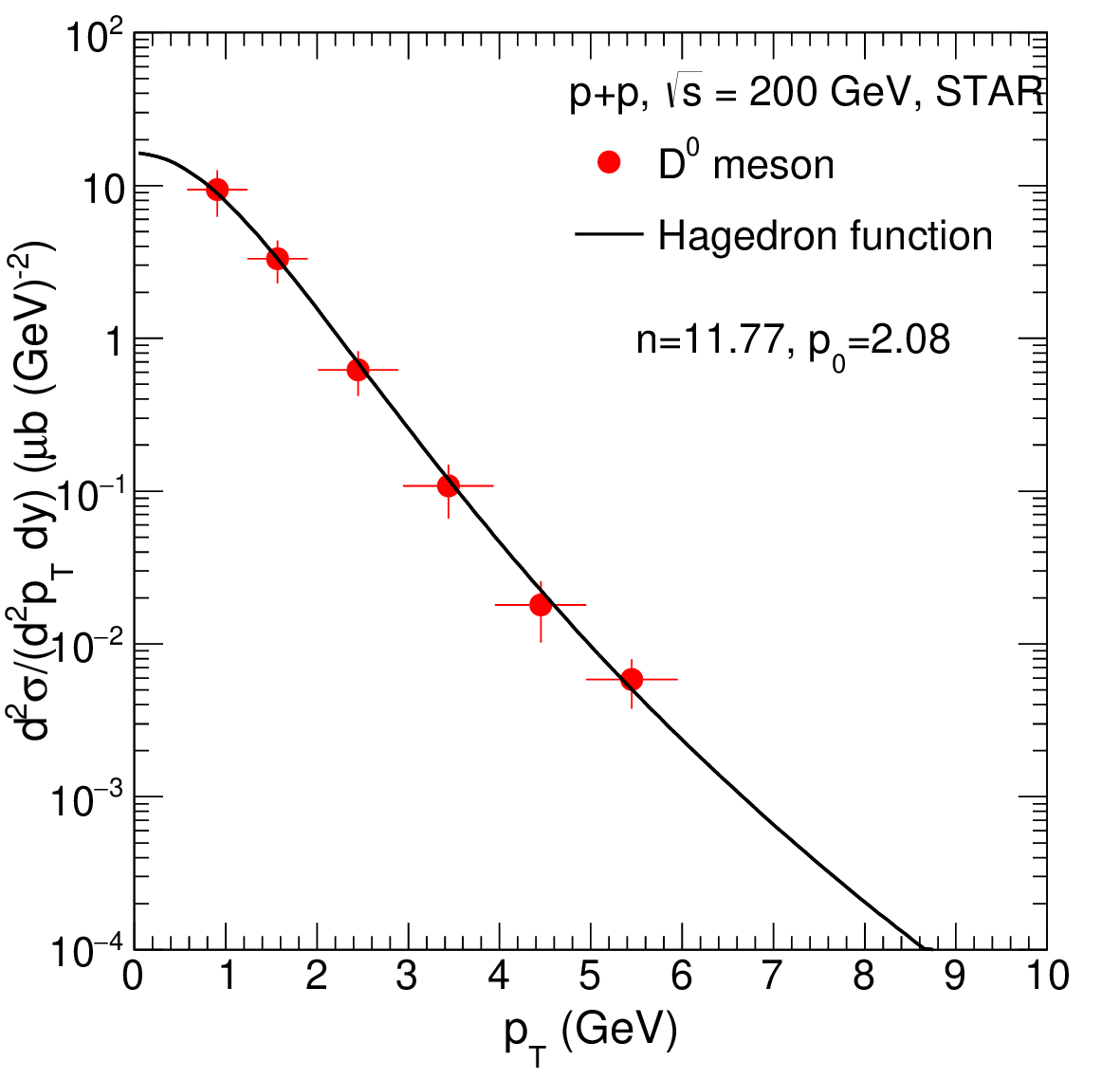}
\includegraphics[width= 0.3\textwidth]{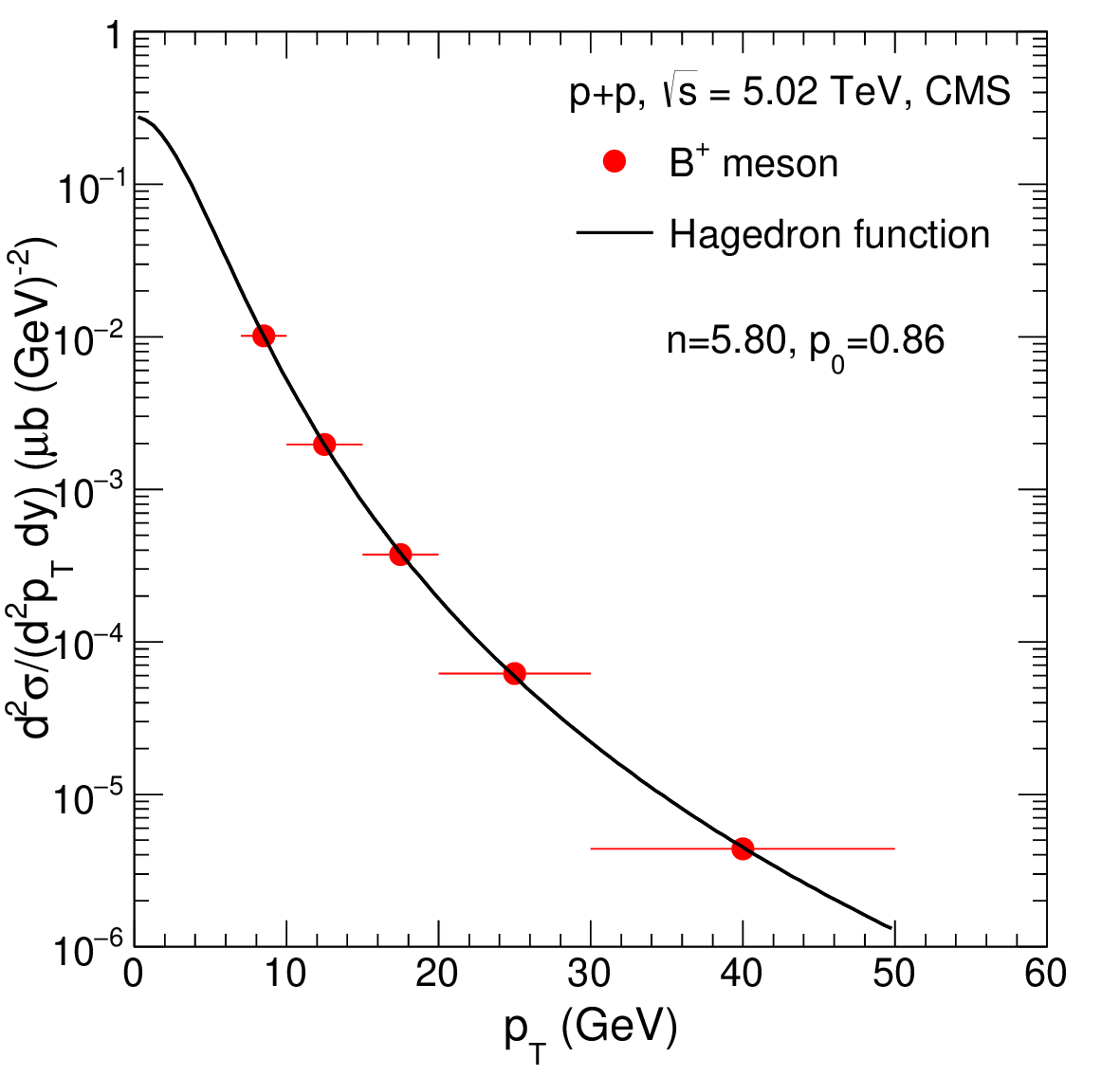}
%\end{center}
\caption{(Colour online) The invariant yields of $D^0$ mesons in p+p collisions
  at $\sqrt{s}=$ 2.76, 5.02 TeV, 200 GeV and $B^+$ mesons in p+p collisions at
$\sqrt{s}=$ 5.02 TeV fitted with Hagedorn function (solid line). 
The data of $D^0$ are adopted from CMS~\cite{CMS_pp1,CMS_pp2}, ALICE~\cite{ALICE_pp1,ALICE_pp2}
and STAR~\cite{STAR_pp} Collaborations. 
The data of $B^+$ are adopted from CMS~\cite{Bplus_pp} Collaboration. The parameter $p_0$ has the unit GeV and $n$ is dimensionless.}
\label{D-LHC}
\end{figure}
%%%%%%%%%%%%%%%%%%%%%%%%%%%%%%%%%%%%%

%%%%%%%%%%%%%%%%%%%%%%%%%%%%%%%%%%
\section{Results of energy loss from experimental data}
%%%%%%%%%%%%%%%%%%%%%%%%%%%%%%%%%
In order to study the energy loss suffered by heavy quarks in QGP, we have fitted
the invariant yields of $D^0$ and $B^+$ mesons for p+p collisions with Hagedorn
function (Eq.~\ref{Hag}) at the RHIC and LHC energies.
The data of invariant yields of $D^0$ and $B^+$ mesons are adopted 
from the CMS~\cite{CMS_pp1,CMS_pp2,Bplus_pp}, ALICE~\cite{ALICE_pp1, ALICE_pp2}
and STAR Collaborations~\cite{STAR_pp}. The CMS and STAR Collaborations have
performed measurement in the rapidity range $|y|<1.0$ whereas ALICE
Collaboration has performed measurement in the rapidity range $|y|<0.5$.
The fit functions along with the data are shown in Fig.~\ref{D-LHC} and the
fit parameters are given in Table~\ref{Table_spectra}. It has been found that
the power $n$ lies between 6--7 for $D^0$ mesons at the LHC energies
($\sqrt{s}=$ 2.76, 5.02 TeV) and is almost 12 for $D^0$ mesons at RHIC
energy ($\sqrt{s}=$ 200 GeV). The power $n$ decreases with 
increasing center of mass energy in agreement with an earlier
work~\cite{Shukla2}. The current study does not include any microscopic dynamics of heavy meson production, 
however our phenomenological study may suggest that the production
of $D^0$ mesons is dominated by leading twist partonic scatterings at the
LHC energies whereas quark-hadron kind of scatterings are more relevant at
the RHIC energy, following the conclusion drawn by similar analysis done for 
light hadron production~\cite{P.khandai}.

%%%%%%%%%%%%%%%%%%%%%%%%%%%%%%%%%%%%%%%%%%%%%%%%%%%%
\begin{figure*}[ht]
\begin{center}
\includegraphics[width= 0.4\textwidth]{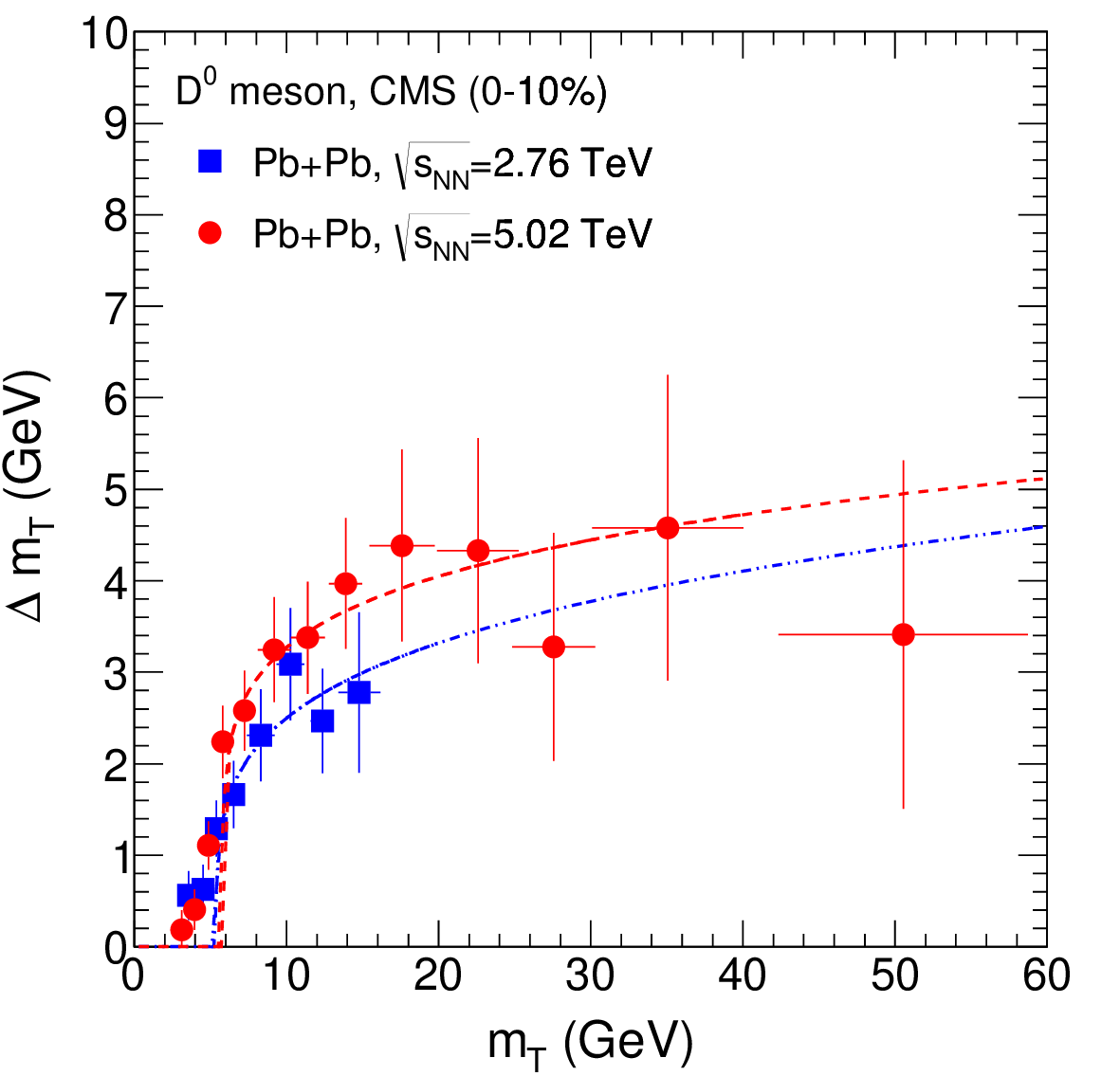}
\includegraphics[width= 0.4\textwidth]{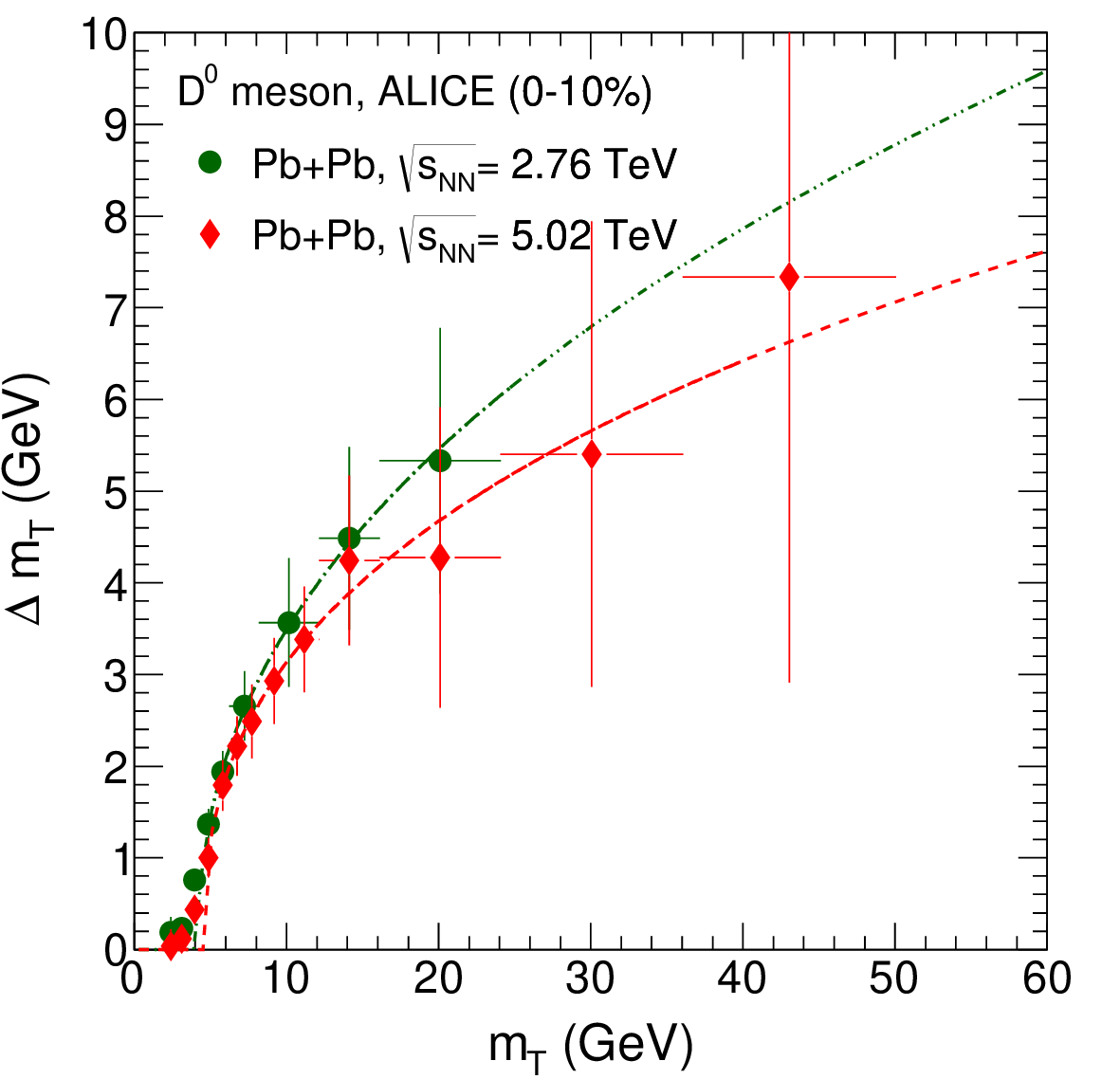}
\includegraphics[width= 0.4\textwidth]{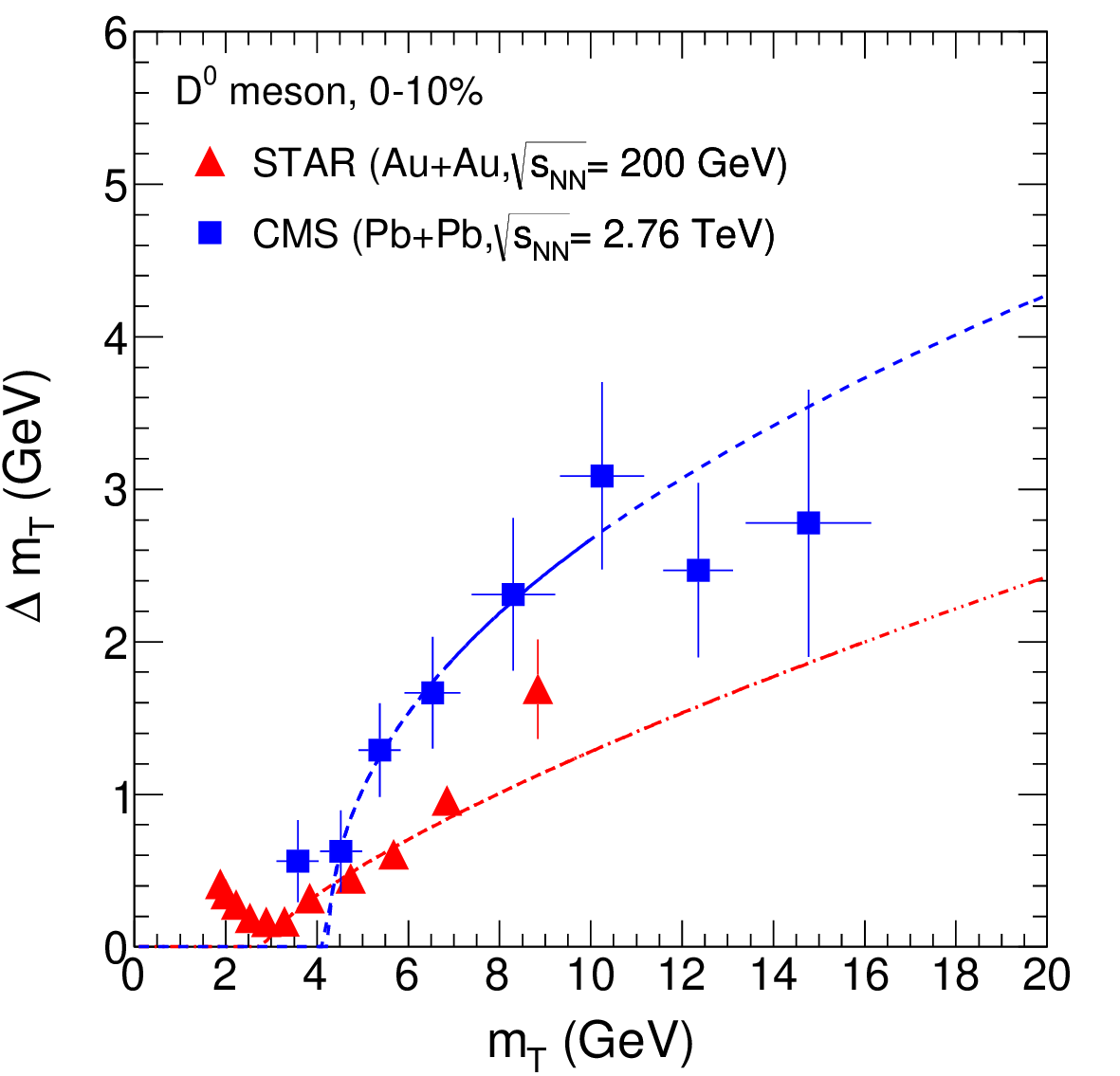}
\includegraphics[width= 0.4\textwidth]{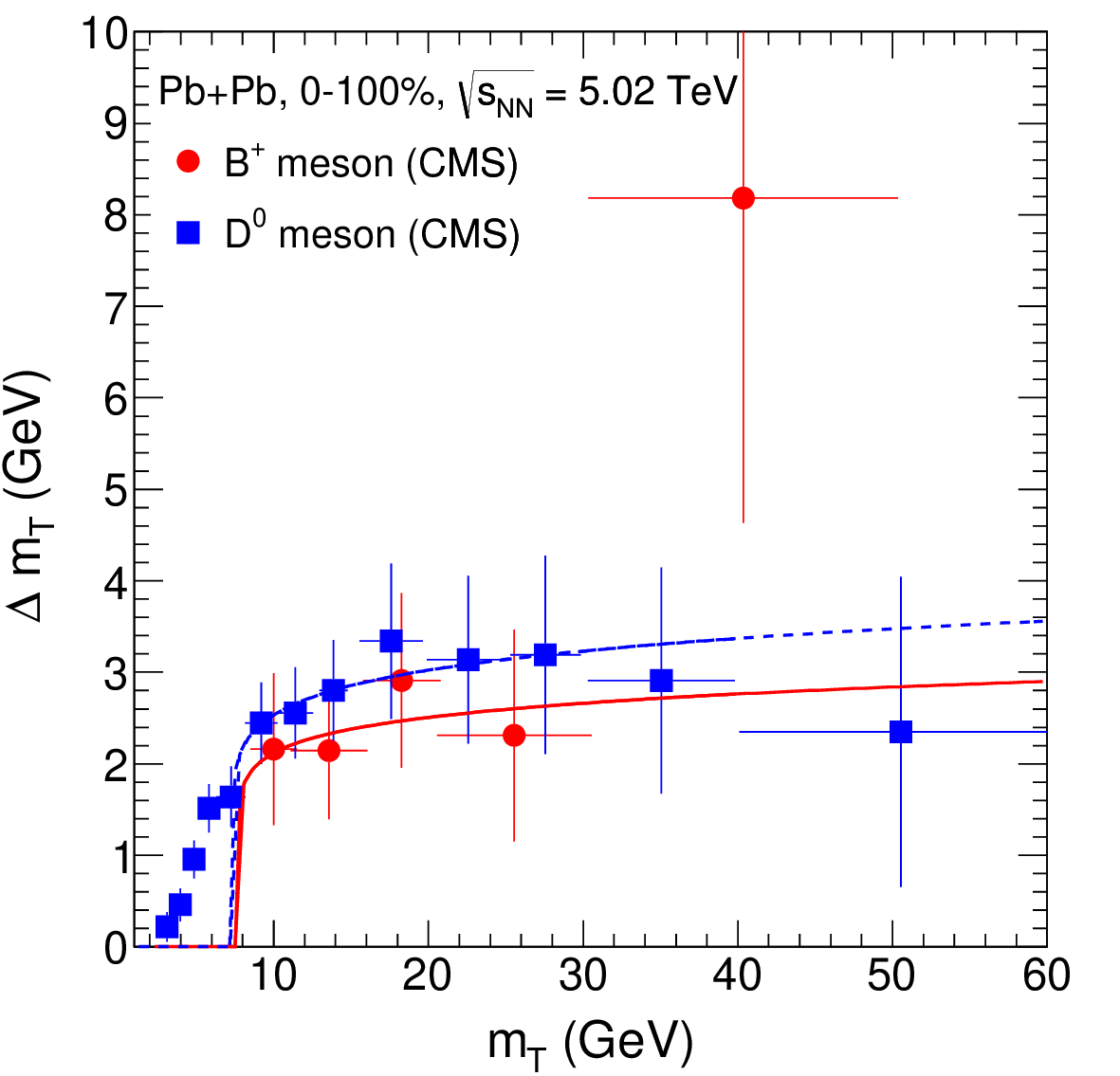}
\end{center}
\caption{(Colour online) The estimated energy loss ($\Delta m_T$) is plotted as a function of transverse mass ($m_T$) of $D^0$ mesons 
in Pb~+~Pb collisions at 
$\sqrt{s_{NN}}=$ 2.76, 5.02 TeV, Au~+Au collisions at $\sqrt{s_{NN}}=$ 200 GeV and $B^+$ mesons in Pb~+~Pb collisions 
at $\sqrt{s_{NN}}=$ 5.02 TeV.  The power law fits are shown by dashed lines.
The data of $R_\textrm{AA}$ of $D^0$ mesons are adopted from CMS~\cite{CMS_pp1, CMS_pp2}, ALICE~\cite{ALICE_pp1, ALICE_pp2} and 
STAR~\cite{STAR_raa} Collaborations. The data of $R_\textrm{AA}$ of $B^+$ mesons are adopted from CMS~\cite{Bplus_pp} Collaboration.}
\label{delmt_meson}
\end{figure*}
%%%%%%%%%%%%%%%%%%%%%%%%%%%%%%%%%%%%%%%%%%%%%%%%%%%%

Next we have estimated the energy loss parameter ($\Delta m_T$) arising
due to the shift in the momentum spectra of final state hadrons ( $D$ or $B$ mesons).
For this purpose, we have inserted the values of the parameters $n$ and $p_0$ 
from Table~\ref{Table_spectra} into Eq.~\ref{delmt}; supplemented with the
$R_{\textrm{AA}}$ of $D^0$ (and $B^+$) mesons measured by the CMS~\cite{CMS_pp1, CMS_pp2}
and ALICE~\cite{ALICE_pp1, ALICE_pp2} Collaborations for Pb+Pb collisions at 
$\sqrt{s_{NN}}=$ 2.76, 5.02 TeV and by the STAR~\cite{STAR_raa} Collaboration for 
Au+Au collisions at $\sqrt{s_{NN}}=$ 200 GeV. Finally $\Delta m_T$  is plotted
against $m_T$ of $D^0$ (and $B^+$) mesons which are depicted in Fig.~\ref{delmt_meson}.
In order to explore the dependence of $\Delta m_T$ on $m_T$, the results are fitted
with a power law:
%%%%%%%%%
\be
\Delta m_T = a_1(m_T-k_1)^{\alpha},
\label{mt-scale}
\ee
%%%%%%%%%
where $a_1$ is normalisation constant, $\alpha$ is the exponent and $k_1$
is the offset transverse mass, is adjusted such that $\Delta m_T$ is positive always.
We have fitted the results for $p_T>$ 4 GeV at the LHC energies.
The limit corresponds to  $m_T\gtrsim$4.5 GeV for $D^0$ and $m_T\gtrsim$ 6 GeV
for $B^+$ mesons. We have empoyed the $\chi^2$ minimisation method
to fit the results. The fit range, fit parameters and $\chi^2$ per degrees of freedom
are listed in Table~\ref{Table_power}. The normalisation parameter ($a_1$)
is proportional to in-medium energy loss; the power ($\alpha$)
could tell us about the energy loss mechanism of heavy quarks in the medium.
The power $\alpha$ close to zero corresponds to coherent regime of energy
loss and $\alpha$ lying between 0.5 and unity corresponds to incoherent or partial
coherent regime of energy loss~\cite{Younus_dk}.
First we have noted that the scaling of $\Delta m_T$ with $m_T$ is clearly
distinguishable for the different scenarios mentioned above.
As seen in Fig.~\ref{delmt_meson} (top-left and top-right panels),
the power laws are found to describe the variation of $\Delta m_T$ with $m_T$
satisfactorily for central (0-10\%)  Pb+Pb collisions at $\sqrt{s_{NN}}=$ 2.76 and 5.02 TeV.
The fit ranges of $m_T$ are: [5:20] GeV at $\sqrt{s_{NN}}=$ 2.76 TeV and [5:40] GeV at $\sqrt{s_{NN}}=$ 5.02 TeV.
The best fit values of ($\alpha$) are 0.24 and 0.45, calculated from the CMS and ALICE measurements at 
$\sqrt{s_{NN}}=$ 2.76 TeV respectively.
The power ($\alpha$) is found to decrease towards higher center of mass energy,
which is about 0.17 for the CMS and about 0.38 for the ALICE at $\sqrt{s_{NN}}=$ 5.02 TeV.
It has been obeserved that ($\alpha$) gradually reduces to smaller values as higher $p_T$ range were included in the fit. 
The observations indicate that the momentum
shift ($\Delta p_T$) of heavy mesons is almost independent of it's transverse
momentum ($p_T$) at large $p_T$. This arises due to coherent energy loss of heavy quarks.
We have noticed slightly different powers ($\alpha$) for the $\Delta m_T$, calculated
for measurements done by ALICE and CMS collaboration. This can be attributed to
the difference in the $R_\textrm{AA}$ mesurements by the two collaborations since the
weightage of high $p_T$ points of ALICE becomes smaller due to larger error bars.

  Fig.~\ref{delmt_meson} (bottom-left panel) shows $\Delta m_T$ for Au+Au
collisions at $\sqrt{s_{NN}}=$ 200 GeV and Pb+Pb collisions at $\sqrt{s_{NN}}=$ 2.76 TeV
for the (0-10\%) centrality of collision. 
 The $p_T$ dependence  of $R_\textrm{AA}$ of $D^0$ meson production exhibit very
similar behaviour, however the magnitudes of $\Delta m_T$ are found quite
different at the two centre of mass energies. The fit range of $m_T$, [3:16] GeV
is used to include the low $p_T$ regions of $R_\textrm{AA}$ measurement. 
The power $\alpha$ is found 0.73 at the RHIC energy and decreases to 0.53 at the
LHC energy. The close proximity of powers signify that the energy loss
mechanism of heavy quarks are similar in the low $p_T$ regions at the LHC
energy and at the RHIC energy. This is the incoherent regime 
of heavy quark energy loss where the $\Delta p_T$ scales as $(p_T)^{1/2}$ and $p_T$.
Fig.~\ref{delmt_meson} (bottom-right panel) shows $\Delta m_T$ for $D^0$ and $B^+$ mesons 
obtained for Pb+Pb collisions at $\sqrt{s_{NN}}=$ 5.02 TeV.  Although the magnitude of
D-meson energy loss is more, they are found to
follow nearly similar power law scaling in range $m_T= [6:40]$ GeV and the power
$\alpha$ comes about 0.1. Thus we can infer that the energy loss of bottom and
charm quark follow same mechanism of energy loss at intermediate and higher $p_T$ regions.

%%%%%%%%%%%%%%%%%%%%%%%%%%%%%%%%%%%%%%
\begin{table*}[!]
\setlength{\arrayrulewidth}{0.12mm}
\setlength{\tabcolsep}{8pt}
\renewcommand{\arraystretch}{0.75}
\begin{tabular}{cccccccc} \hline
\multicolumn{1}{c}{$\sqrt{s_{NN}}$} & 
\multicolumn{1}{c}{Experiment} &
\multicolumn{1}{c}{Meson} &
\multicolumn{1}{c}{Fit Range} &
\multicolumn{1}{c}{Norm} &
\multicolumn{1}{c}{Power} & 
\multicolumn{1}{c}{$k_1$} &
\multicolumn{1}{c}{$\chi^2$} \\
(TeV) &  &  &$m_T$ (GeV)& ($a_1$) & ($\alpha$) &(GeV)& /NDF\\
\hline\hline
2.76 & CMS(0-10\%) & $D^0$ & [5.0,20.0] & 1.71$\pm$0.38 & 0.24$\pm$0.14 & 5.36$\pm$0.01 & 0.401 \\
5.02 & CMS(0-10\%) & $D^0$ & [5.0,40.0] & 2.45$\pm$0.42 & 0.17$\pm$0.08 & 5.80$\pm$0.08 & 0.208 \\
2.76 & ALICE(0-10\%) & $D^0$ & [5.0,20.0] & 1.59$\pm$0.80 & 0.45$\pm$0.07 & 4.17$\pm$0.70 & 0.009 \\
5.02 & ALICE(0-10\%) & $D^0$ & [5.0,40.0]  & 1.63$\pm$0.98 & 0.38$\pm$0.06 & 4.57$\pm$0.70 & 0.043 \\
0.200 & STAR(0-10\%) & $D^0$ & [3.0,16.0] & 0.30$\pm$0.05 & 0.73$\pm$0.14 & 2.83$\pm$0.20 & 3.536 \\
2.76 & CMS(0-10\%) & $D^0$ & [3.0,16.0] & 0.90$\pm$0.21 & 0.53$\pm$0.13 & 3.57$\pm$0.004 & 0.487 \\
5.02 & CMS(0-100\%) & $B^+$ & [6.0,40.0] & 1.95$\pm$0.97 & 0.10$\pm$0.06 & 7.72$\pm$0.85 & 0.338 \\
5.02 & CMS(0-100\%) & $D^0$ & [6.0,40.0] & 2.24$\pm$0.62 & 0.11$\pm$0.08 & 7.24$\pm$0.41 & 0.070 \\
\hline
%\hline
\end{tabular}
\caption{The parameters of power law (Eq.~\ref{mt-scale}) obtained by fitting $\Delta m_T$
 as a functiom of $m_T$ for D and B mesons at RHIC and LHC. 
The chi-square per degrees of freedom has also shown for each case.}
\label{Table_power}
%\end{widetext}
\end{table*}
%%%%%%%%%%%%%%%%%%%%%%%%%%%%%%%%%%%%%%%%%%%%%%%%%%%%%

%%%%%%%%%%%%%%%%%%%%%%%%%%%%%%%%%%%%%
\subsection{\bf Centrality dependence of $\Delta m_T$}
%%%%%%%%%%%%%%%%%%%%%%%%%%%%%%%%%%%%%

Figure~\ref{mt-centrality} shows energy loss ($\Delta m_T$) as a function of
transverse mass ($m_T$) of $D^0$ mesons for different centralities of
Pb~+~Pb collisions at $\sqrt{s_{NN}}=$ 5.02 TeV.
  The data of $R_\textrm{AA}$ of $D^0$ mesons are adopted from 
ALICE~\cite{ALICE_pp2} Collaborations.
In order to compare the fit parameters we have chosen a common fit range [6:30] GeV
for all the three centralities of collision. It is found that the normalisation
parameter ($a_1$) gradually increases from the perpheral to central collisions
as the system size (hence pathlength) increases with the number of
participant nucleons ($N_{part}$)~\cite{ALICE_Npart}.
The normalisation parameter $a_1$ roughly scales as $\sqrt{N_{part}}$ which gives the system size.
The power ($\alpha$) remains almost same for the three centralities while considering the error in 
the fitting. The energy loss of the charm quark is directly depends on the the temperature of the medium.
Now the hydrodynamics studies suggest that the average temperature of the medium increases with center of mass 
energy of collisions and changes very little with centrality collision for a given center of mass energy~\cite{kolb1, kolb2, pingal-fcc}.
Thus we may conclude that energy loss mechanism chiefly depends on the
collision energy and has a weak dependence on centrality of collision.
%%%%%%%%%%%%%%%%%%%%%%%%%%%%%%%%%%%%
\begin{table}[ht]
%\begin{center}
\begin{tabular}{| c | c | c | c | c | c |} 
\hline
 Centrality   &    Norm   &    Power     &  Offset mass   & $\chi^2/NDF$  & $ N_{part}$      \\ 
 class ($\%$)  &  ($a_1$)     & ($\alpha$) &   ($k_1$ in GeV)    &           &                   \\ \hline
   0-10        & 1.64$\pm$0.01 &  0.38$\pm$0.09 &  4.65$\pm$0.67  &  0.06    &    359.0           \\ \hline 
   30-50       & 0.79$\pm$0.05 &  0.40$\pm$0.05  &  3.46$\pm$0.58  & 0.14     &   118.5            \\ \hline 
   60-80       & 0.45$\pm$0.07 &  0.45$\pm$0.02   &  5.99$\pm$0.52   & 0.19    &  23.0              \\ \hline 
\end{tabular}
%\end{center}
\caption{The parameters referred in Eq.~\ref{mt-scale} for three different 
  centralities of Pb~+~Pb collisions at $\sqrt{s_{NN}}=$ 5.02 TeV.
  The fit range is [6:30] GeV for all centralities.}
\label{mt-centrality}
\end{table}
%%%%%%%%%%%%%%%%%%%%%%%%%%%%%%%%%%%%%%%%

%%%%%%%%%%%%%%%%%%%%%%%%%%%%%%%%%%%
\begin{figure*}[t]
\begin{center}
\includegraphics[width= 0.5\textwidth]{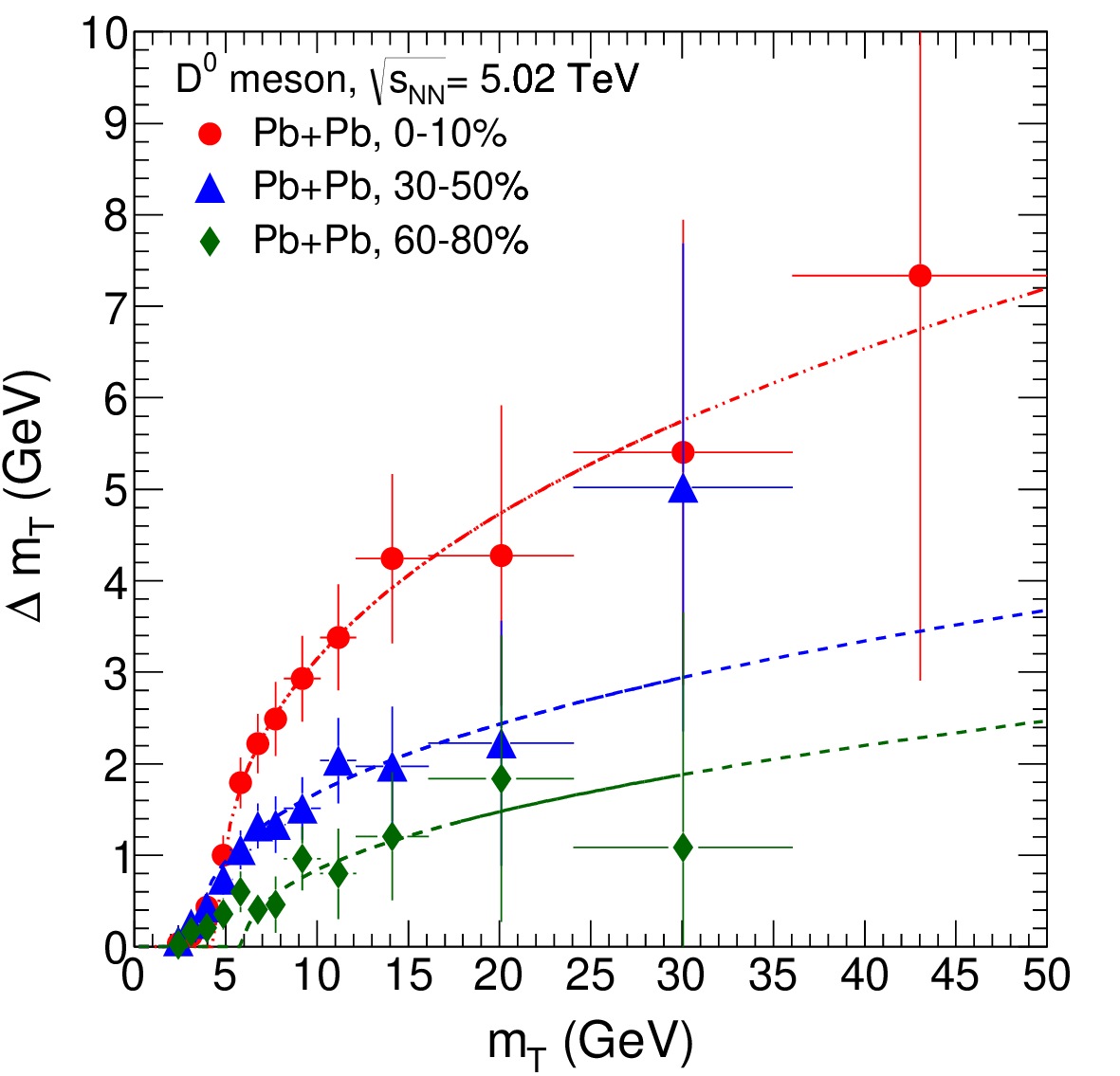}
\end{center}
\caption{(Colour online) Energy loss ($\Delta m_T$) as a function of
transverse mass ($m_T$) of $D^0$ mesons for different centralities of
Pb~+~Pb collisions at $\sqrt{s_{NN}}=$ 5.02 TeV.
  The data of $R_\textrm{AA}$ of $D^0$ mesons are adopted from 
ALICE~\cite{ALICE_pp2} Collaborations.}
\label{delmt_centrality}
\end{figure*}
%%%%%%%%%%%%%%%%%%%%%%%%%%%%%%%%%%%%%%%%

%%%%%%%%%%%%%%%%%%%%%%%%%%%%%%%%%%%%%%%%%%%%
\section{Energy loss from theoretical models}
In order to get more insight of our empirical analysis, we have calculated
transverse energy loss ($\Delta E_T$) of a heavy quark inside QGP at the LHC
and RHIC energies. The transverse energy of a heavy quark is defined as:
$E_T= \sqrt{p_T^2~+~m_Q^2}$, where $p_T$ is the transverse momentum and $m_Q$ 
is the mass of the heavy quark. The heavy quarks lose energy through collisions
with medium partons and radiation of gluons. Thus, $\Delta E_T$ is quite relevant
to the observable $\Delta m_T$, which we have proposed as an alternative measure
of energy loss of heavy quarks in QGP. We have not included any charm quark
hadronization mechanism in the present work. Several formalisms have been proposed over
the past few decades for the estimation of collisional and radiative energy loss
of heavy quarks~\cite{BT-coll, PP-coll, DGLV, DGLV2, XDZR, ASW, Abir_Upoff}.
In addition, there are energy loss models developed in recent times, have
described the nuclear suppression and azimuthal anisotropy of D mesons very
well in different regions of $p_T$~\cite{BAMPS,DUKE,TAMU, MCEPOS, CUJET, POWLANG, PHSD,SKD}.
In the present study, we have followed Djordjevic, Gyulassy, Levai, and Vitev (DGLV)
formalism using opacity expansion~\cite{DGLV,DGLV2}, the treatment of
Xiang, Ding, Zhou, and Rohrich (XDZR) using light cone path integral
approach~\cite{XDZR}, and the generalised dead cone approach by the authors in
Ref.~\cite{Kapil_NPA} for the radiative energy loss. The soft gluon emission
from a heavy quark is suppressed in comparison to that from a light quark due
to it's large mass. This is known commonly as dead cone effect~\cite{Dead-cone}. 
Later it has been shown that the effect of dead-cone diminishes when the energy
of heavy quark is large compared to it's mass~\cite{Abir_Upoff}. 
The formalism introduced by Abir, Jamil, Mustafa and Srivasatava (AJMS) was
found to describe satisfactorily the nuclear modification factor ($R_{\textrm{AA}}$)
of D mesons at RHIC and the LHC energies~\cite{Abir_Jamil}.
The AJMS formalism was further modified by the authors in Ref~\cite{Kapil_NPA}
which is referred as \textit{Corrected} AJMS which we have incorporated 
in this work. The treatment of Peigne and Peshier (PP)~\cite{PP-coll}
has been adopted for the calculation of collisional energy loss.
The mathematical formulas of all the formalisms are given in the appendix.

%%%%%%%%%%%%%%%%%%%%%%%%%%%%%%%%%%%%%%%%%%%%%%%%%%%
\begin{figure*}[]
\begin{center}
\includegraphics[width= 0.45\textwidth]{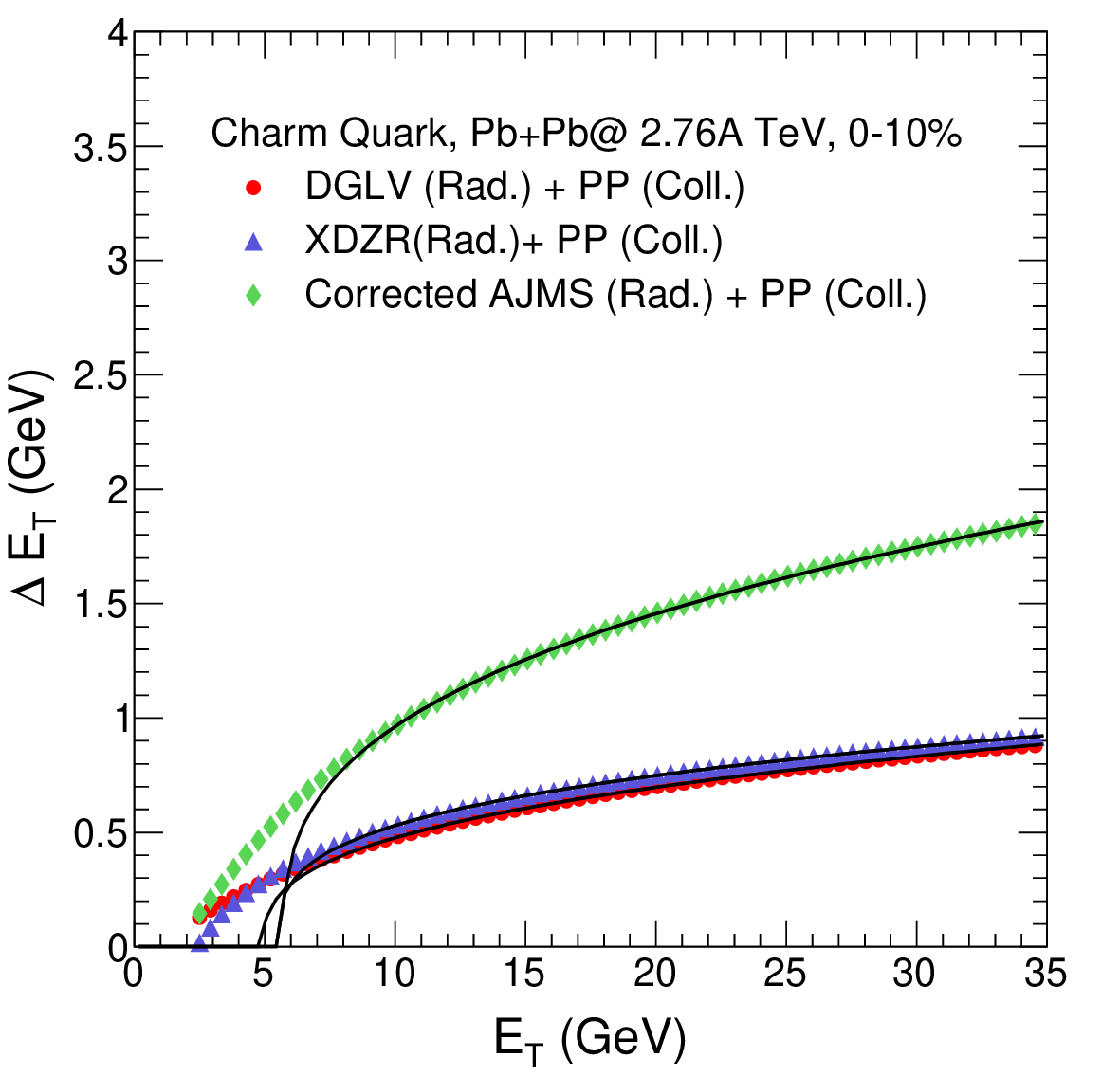}
\includegraphics[width= 0.45\textwidth]{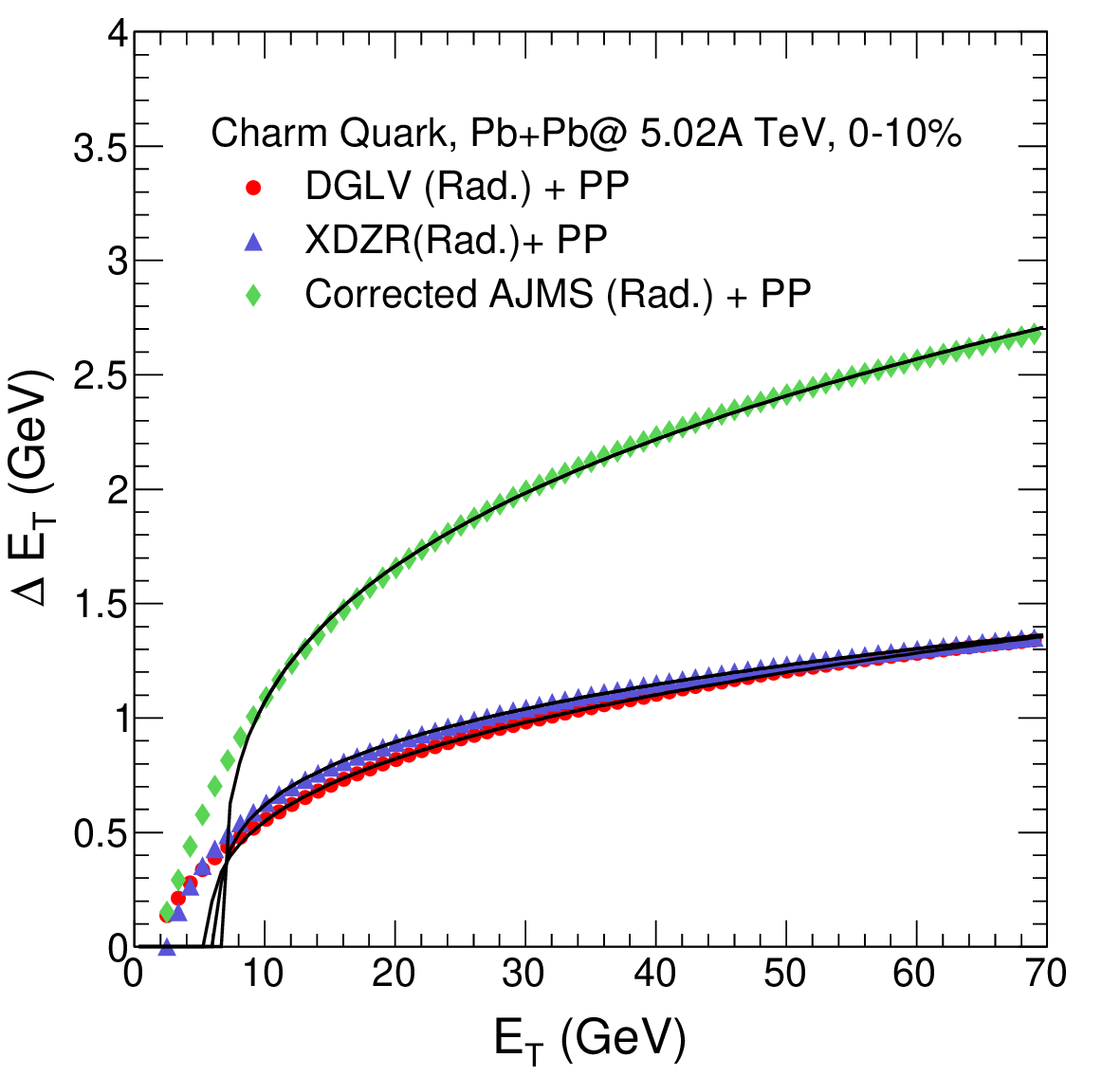}
\includegraphics[width= 0.45\textwidth]{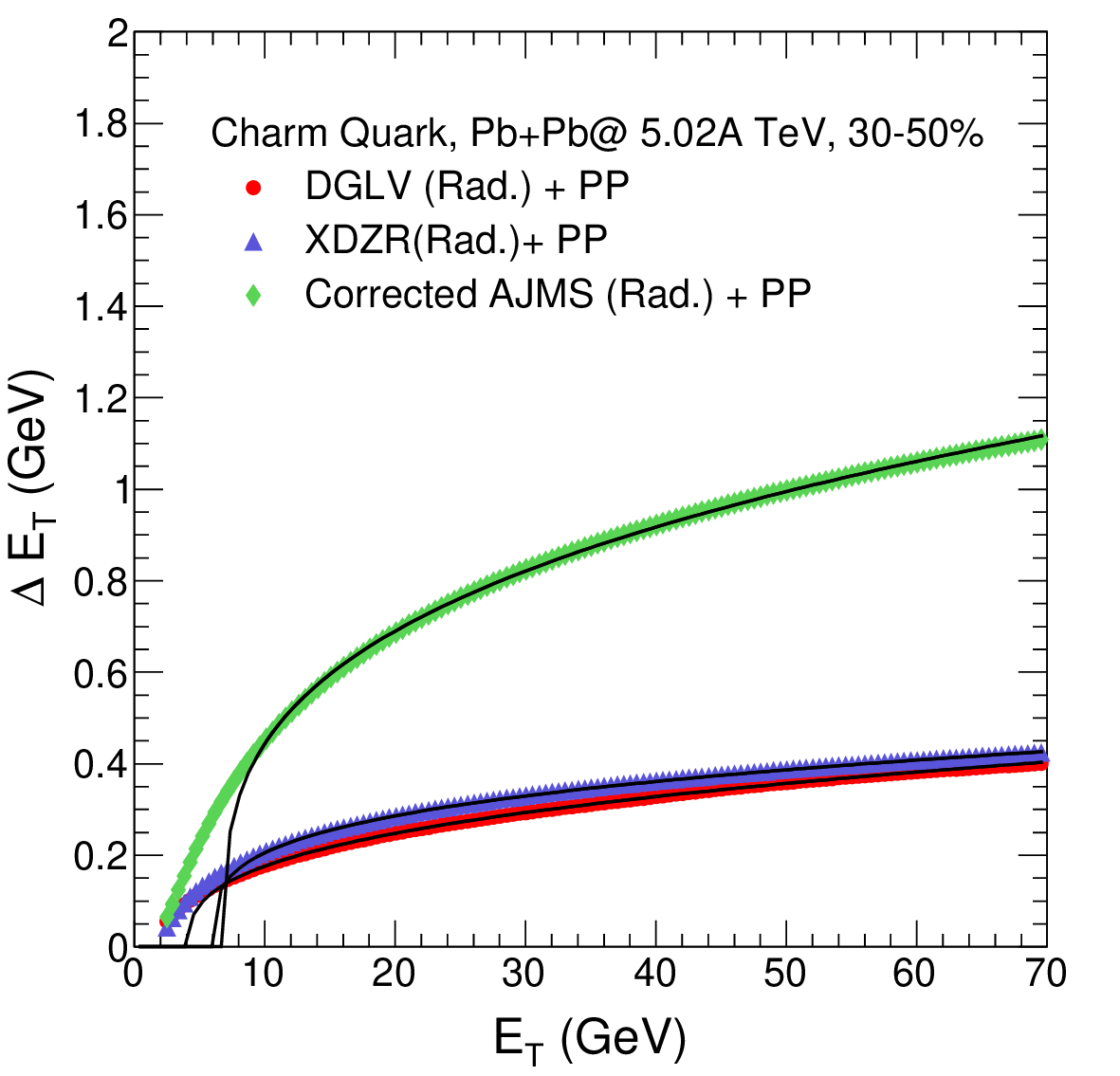}
\includegraphics[width= 0.45\textwidth]{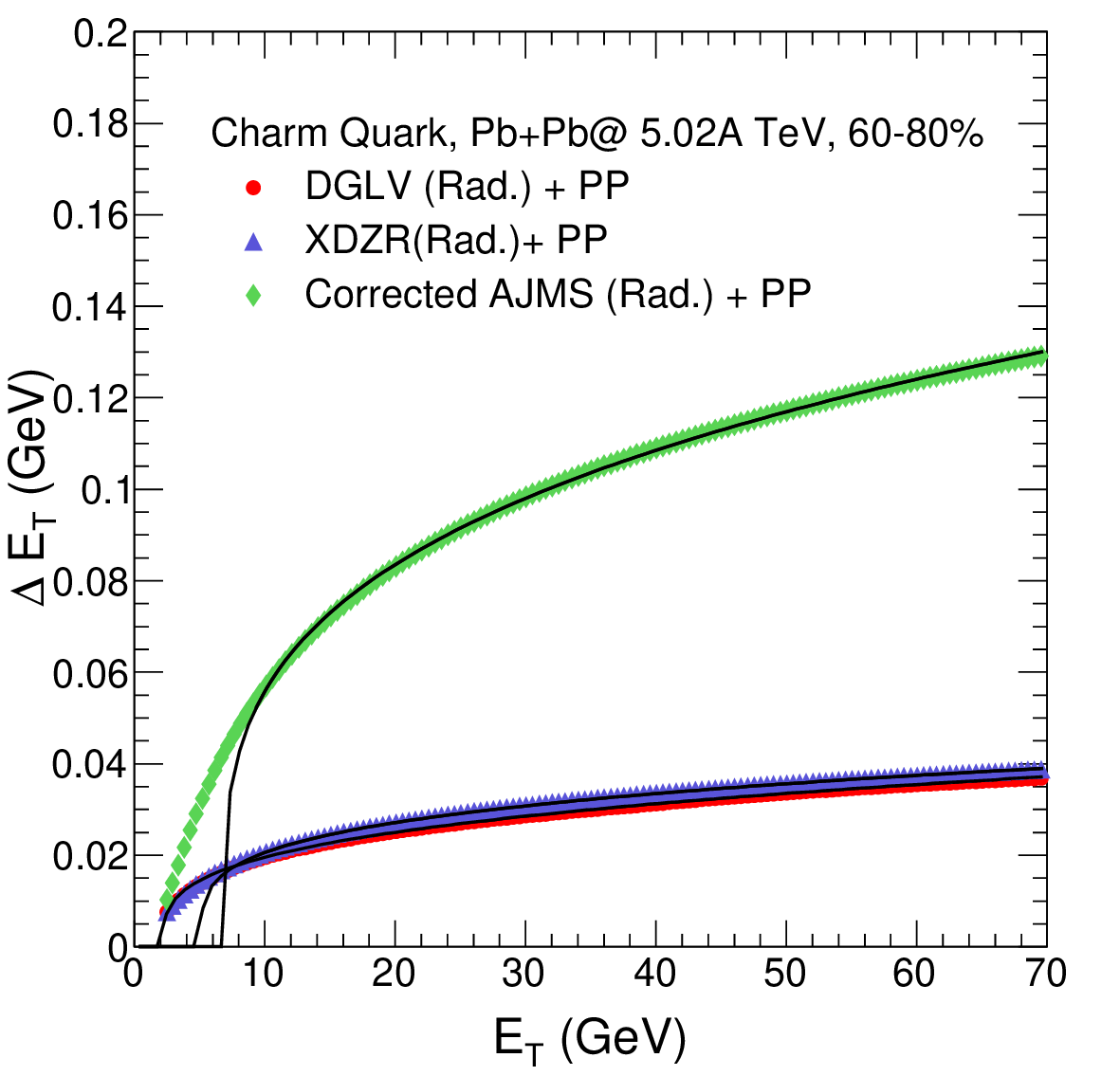}
\end{center}
\caption{(Colour online) The Collisional+Radiative energy loss suffered by
  a charm quark is plotted as function of transverse energy ($E_T$) in Pb+Pb collisions at $\sqrt{s_{NN}}=$2.76 TeV (upper left) and 5.02 TeV (upper right) for 0-10\% centrality. 
  Lower panel shows the results for 30-50\% (left) and 60-80\% (right) centralities in Pb+Pb collisions at $\sqrt{s_{NN}}=$5.02 TeV.
}
\label{delET_HQ}
\end{figure*}
%%%%%%%%%%%%%%%%%%%%%%%%%%%%%%%%%%%%%

\subsection{The evolution of QGP}
The energy loss of the quarks depends upon the path length traversed
inside the plasma and the temperature of the plasma in addition to the energy of
the heavy quark. The collisional and radiative energy losses of
a heavy quark are calculated as a function of temperature which is then averaged
over the time evolution of the plasma. The temporal evolution of temperature of the
thermalised medium is calculated using 2+1 dimensional 
ideal hydrodynamics simulation~\cite{kolb1, kolb2, pingal-fcc}.
The critical temperature of quark-hadron phase transition is considered at 170 MeV,
following the lattice QCD based EoS~\cite{lattice}. The hydrodynamics simulation
has been used to estimate the thermal photon production for Pb+Pb collisions 
at $\sqrt{s_{NN}}=$ 2.76 TeV and 5.02 TeV at different centralities of
collision~\cite{pingal-fcc}. Assuming the heavy quark is produced at a point P$(r,\phi)$ in
the reaction plane and it travels a distance $l(r,\phi)$ in azimuthal direction
$\phi$ with respect to reaction plane, then the average path length for an impact parameter ($b$)
is given by~\cite{S_dk1}:\\
\begin{equation}
  <L> = \frac{\int_0^R \! \int_0^{2\pi} l(r,\phi) T_{AB}(\vec{b})\, rdr\,d\phi}
    {\int_0^R \! \int_0^{2\pi} T_{AB}(\vec{b})\, rdr\,d\phi}    ,\\
\end{equation}
where $T_{AB}(\vec{b})= t_A(\vec{r}+\vec{b}/2)t_B(\vec{r}-\vec{b}/2)$ represents the number density
of nucleons in the transverse plane, $t_A$ and $t_B$ are the thickness functions
of the colliding nuclei A and B respectively.
Let the velocity of the heavy quark is $v_T(= p_T/ E_T)$ and
$\tau_c$ is the proper time which corresponds to the critical temperature $T_c$
of quark-hadron phase transition.
Then the effective path length of the heavy quark is $< L>_{eff}= {\rm min}[<L>, v_T \times \tau_c]$.
The procedure was adopted in earlier works~\cite{DGLV2,Jamil}.
The effective paramaters of the system considered in the calculations
are enlisted in Table~\ref{qgp_evolution}.
%%%%%%%%%%%%%%%%%%%%%%%%%%%%%%%%%%%%
\begin{table}[ht]
%\begin{center}
\begin{tabular}{| c | c | c | c | c | c |} 
\hline
$\sqrt{s_{NN}}$ & Centrality & $<b>$ &  $N_{\rm{part}}$ & $<L>$   & $\tau_{c}$  \\ 
   (TeV)       & class ($\%$)     & (fm)  &             & (fm)   & (fm) \\ \hline\hline
   2.76        & 0-10     & 3.30       &  351.15        &  4.890  & 7.525      \\ \hline 
   5.02        & 0-10     & 3.36        & 352.13      &  4.867  & 8.155       \\ \hline 
   5.02        & 30-50    & 9.95       &  93.50         &  2.124   & 5.530       \\ \hline
   5.02        & 60-80    & 12.80      &   21.30       &  0.415  & 1.890       \\ \hline 
 %  2.76        & 30-50    & 9.90      &   88.57        &   2.150  & 5.530       \\ \hline 
\end{tabular}
%\end{center}
\caption{The effective paramaters of the system considered in the calculations.}
\label{qgp_evolution}
\end{table}
%%%%%%%%%%%%%%%%%%%%%%%%%%%%%%%%%%%%%%%%

\subsection{Results}
We have calculated the transverse energy loss ($\Delta E_T$) of a charm
quark for the central (0-10\%) Pb+Pb collisions  at $\sqrt{s_{NN}}=$ 2.76 TeV
and for the three centralities 0-10\%, 30-50\%, 60-80\% of Pb+Pb collisions at $\sqrt{s_{NN}}=$ 5.02 TeV. 
The rapidity ($y$) for all collision centralities
are taken as $y=1.0$. The results of combined collisional and radiative energy
losses are depicted in Fig.~\ref{delET_HQ}. Let us discuss significant features
of the results. It can be noted that DGLV+PP and XDZR+PP formalisms yield similar
amount of energy losses of a charm quark for a given system except in a small
region ($E_T<5$ GeV).
The results have come as per expectation because the above two 
approaches have considered similar kind of assumption in the calculations e.g,
static Debye screened scattering centres and LPM suppression effect.
On the other side, the energy loss obtained from \textit{Corrected}
AJMS+PP formalism is found higher than that by the other two formalisms,
by about a factor of two. This can be attributed to the fact that the kinematic cuts are less
restricted thus a broader range of gluon emission angle is utilised in
\textit{Corrected} AJMS approach. 

Next, we would like to examine that whether a power law, analogous to Eq.~\ref{mt-scale},
could describe the variation of $\Delta E_T$ as a function of $E_T$ as well. 
Thus we write the formula:
%%%%%%%%%%%
\be
\Delta E_T = a_2\,(E_T-k_2)^{\beta},
\label{et-scale}
\ee
%%%%%%%%%%%%
where $a_2$ is the normalisation constant and $\beta$ is the exponent. The offset
transverse mass $k_2$, is adjusted in such a way that $\Delta E_T$ is positive always.
We have fitted the results using $\chi^2$ minimisation technique in range $E_T=[9.0,35.0]$ GeV for Pb+Pb collisions at
$\sqrt{s_{NN}}=$2.76 TeV and $E_T=[9.0,70.0]$ GeV for Pb+Pb collisions 
at $\sqrt{s_{NN}}=$5.02 TeV. The choice of fit 
range is based upon the scenario that a $D^0$ meson retains about $60\%$ of
transverse energy when it is fragmented from a charm quark. The idea was suggested
by a global analysis of charm quark fragmentation measurement 
in proton-proton collisions at the LHC energy~\cite{Charm-fragmentation}.
The best fit values of the exponent $\beta$ for three energy loss formalisms
are given in Table~\ref{theory_fit}. The $\chi^2 / NDF$ value lies between $10^{-4}$ to $10^{-5}$ for the different
energy loss scenarios.
%%%%%%%%%%%%%%%%%%%%%%%%%%%%%
\begin{table}[ht]
%\begin{center}
\begin{tabular}{| c | c | c | c | c |} 
\hline
$\sqrt{s_{NN}}$ & Centrality & $DGLV$       &  $XDZR$      &$Corr. AJMS$   \\ 
   (TeV)       & class ($\%$)& $(\beta)$   &    $(\beta)$   & $(\beta)$     \\ \hline\hline 
   2.76        & 0-10     & 0.35$\pm$0.002        &  0.29$\pm$0.001         & 0.34$\pm$0.002    \\ \hline 
%   2.76        & 30-50    & 0.40       &  0.39         & 0.49      \\ \hline 
  5.02        & 0-10     & 0.32$\pm$0.002       &  0.27$\pm$0.001        & 0.30$\pm$0.002      \\ \hline 
   5.02        & 30-50    & 0.34$\pm$0.001       &  0.26$\pm$0.001         & 0.30$\pm$0.002        \\ \hline 
    5.02        & 60-80    & 0.28$\pm$0.0001       &  0.24$\pm$0.0001         & 0.28$\pm$0.0001        \\ \hline
\end{tabular}
%\end{center}
\caption{The best fit values of the exponent $\beta$ for three energy loss formalisms
  in case of Pb+Pb collisions at $\sqrt{s_{NN}}=$ 2.76 and 5.02 TeV.}
\label{theory_fit}
\end{table}
%%%%%%%%%%%%%%%%%%%%%%%%
It is observed that the values of $\beta$ obtained from $DGLV$ and $Corr. AJMS$ energy
loss models are quite similar while the $XDZR$ model yields slightly lower value
of $\beta$. Next we consider any one formalism say; $DGLV$, we have found that
$\beta$ is about 0.35 at $\sqrt{s_{NN}}=$2.76 TeV and $\beta$ is about 0.32
at $\sqrt{s_{NN}}=$5.02 TeV. Thus $\beta$ is decreasing towards higher centre of mass energy, 
a similar feature was observed for the exponent $\alpha$ from $\Delta m_T$ scaling.
We have found that the value of $\beta$ for the three collision centralities at $\sqrt{s_{NN}}=$5.02 TeV
are nearly similar which support our earlier conclusion derived from the centrality dependence $\Delta m_T$ at 
sub-section 3.1.
It may also be noted that the values of exponent $\alpha$ empirically obtained from
$\Delta m_T$ scaling (Table~\ref{Table_power}) are quite similar in magnitude
to the values of $\beta$ for the same centrality and collision energy.  
From this preliminary calculation we may infer that the parameter $\Delta m_T$ is
quite relevant to the in-medium heavy quark energy loss and a more dedicated
calculation of heavy quark energy loss supplemented with heavy quark hadronization could firmly establish the correspondence.  
%%%%%%%%%%%%%%%%%%%%%%%%%%%%%%%%%%%%%%%%%%%%

\section{Summary}
In this article, we have proposed an analytical and empirical approach to
estimate the energy loss of heavy quarks in the hot and dense partonic medium
created in the heavy ion collisions at the RHIC and the LHC energies. For this purpose, we have parameterised the invariant
yields of heavy mesons($D^0$ and $B^+$) measured in proton-proton collisions using Hagedorn
function. The parameters of Hagedorn function and the measured nuclear modification
factor of heavy mesons are then utilised to estimate the energy loss of
heavy quarks through the parameter $\Delta m_T$. The parameter $\Delta m_T$ is found to scale with
the transverse mass ($m_T$) of heavy mesons through a power law both at the RHIC and LHC energies. It is also found 
that the scaling of $\Delta m_T$ vs. $m_T$ is clearly discernible
for the two different center of mass energies of the LHC experiment. The exponent of scaling ($\alpha$) 
is found to be smaller at higher center of mass energies which signify the coherent regime of quark energy loss.
We have also calculated the transverse energy loss of charm quarks for central and
mid-central Pb+Pb collisions at $\sqrt{s_{NN}}=$ 2.76 TeV and 5.02 TeV.
We have followed three well known formalisms of radiative energy loss with the most
trusted collisional energy loss formalism. The evolution of the QGP medium is
governed by (2+1) dimensional ideal hydrodynamics. The total
transverse energy loss of charm quarks is also found to scale with the power of
transverse mass in a similar fashion. The best fit values of powers are of
similar magnitude for the two scenarios when we overlook the uncertainties
(which chiefly arise due to large uncertainties in the experimental data itself).
Thus we have advocated that the energy loss $\Delta m_T$ could be an alternative measure of heavy quark
energy loss in QGP medium. More precise experimental data of heavy mesons in near
future would be helpful for the understanding of scaling behaviour of $\Delta m_T$
over wide region of $m_T$ ($p_T$).

\section*{ACKNOWLEDGEMENTS}
The authors are thankful to Kapil Saraswat for many stimulating discussions regarding
the work. We are also thankful to Pingal Das Gupta for his useful aid on ideal hydrodynamics.
Sudipan De acknowledges financial support from the DST INSPIRE Faculty research grant
(IFA18-PH220), India.

%\begin{appendices}

%%%%%%%%%%%%%%%%%%%%%%%%%%%%%%%%%%%%%%%%%%%%%%%%%%%%%%%%%
\section*{Appendix I: Collisional Energy Loss}\label{secA1}
\subsection*{1. Peigne and Peshier \cite{PP-coll}}
\be
\frac{dE}{dx} = \frac{8\pi\alpha^{2}_{s}T^{2}}{3}\left(1 + \frac{N_{f}}{6}\right)log\left(2^{\frac{N_f}{2\left(6 + N_f\right)}}0.92\frac{\sqrt{ET}}{m_g}\right)\nn.
\ee

%%%%%%%%%%%%%%%%%%%%%%%%%%%%%%%%%%%%%%%%%%%%%%%%%%%%%%%%%%%%%%
\section*{Appendix II: Radiative Energy Loss}
\subsection*{1. Djordjevic, Gyulassy, Levai, and Vitev \cite{DGLV,DGLV2}}
\bea
\Delta E = \frac{c_F\alpha_s}{\pi}\frac{EL}{\lambda_g}\int_{\frac{m_g}{E+p}}^{1-\frac{M}{E+p}}\,dx\nn\\
\int_{0}^{\infty}\frac{4\mu^2q^3dq}{\left(\frac{4Ex}{L}\right)^2 + \left(q^2 + \beta^2\right)^2}\nn.
\left(AlogB + C\right)\nn ,
\eea
where\\

$\beta^2 = m_g^2(1-x) + M^2x^2$,\\

$\frac{1}{\lambda_g}=\rho_g\sigma_{gg} + \rho_q\sigma_{qg}$,\\

$\sigma_{gg} = \frac{9\pi\alpha_s^2}{2\mu^2}$, \,\,
$\sigma_{qg} = \frac{4}{9}\sigma_{gg}$,\\

$\rho_g = 16T^3\frac{1.202}{\pi^2}$, \,\,
$\rho_q= 9N_fT^3\frac{1.202}{\pi^2}$.\\
\be
A = \frac{2\beta^2}{f_\beta^3}\left(\beta^2 + q^2\right),\nn\\
\ee
\be
B = \frac{\left(\beta^2 + K\right)\left(\beta^2Q_\mu^- + Q_\mu^+Q_\mu^+ + Q_\mu^+f_\beta \right)}{\beta^2\left(\beta^2\left(Q_\mu^- - K\right) - Q_\mu^-K + Q_\mu^+Q_\mu^+ + f_\beta f_\mu  \right)},\nn\\
\ee
\bea
C = \frac{1}{2q^2f_\beta^2f_\mu}[\beta^2\mu^2\left(2q^2 - \mu^2\right) + \beta^2\left(\beta^2 - \mu^2\right)K\nn\\
+ Q_\mu^+\left(\beta^4 - 2q^2Q_\mu^+\right)\nn\\
+ f_\mu \left(\beta^2\left(- \beta^2 - 3q^2 + \mu^2\right) + 2q^2Q_\mu^+\right) + 3\beta^2q^2Q_k^-],\nn
\eea

$K = \left(2px\left(1-x\right)\right)^2$,\nn\\

$Q_\mu^\pm = q^2 \pm \mu^2$, \,\,
$Q_k^\pm = q^2 \pm K$,\\

$f_\beta = f(\beta, Q_\mu^-, Q_\mu^+)$, \,\,
$f_\mu = f(\mu, Q_k^+, Q_k^-)$, \\

$f(x,y,z) = \sqrt{x^4+2x^2y+z^2}$.
%%%%%%%%%%%%%%%%%%%%%%%%%%%%%%%%%%%%%%%%%%%%%%%%%
\subsection*{2. Xiang, Ding, Zhou, and Rohrich \cite{XDZR} }
\bea
\Delta E = \frac{\alpha_s c_F}{4}\frac{L^2\mu^2}{\lambda_g}  \bigg[ &  log\frac{E}{\omega_{cr}} + \frac{m_g^2L}{3\pi\omega_{cr}}\bigg(1- \frac{\omega_cr}{E}log\frac{E^2}{2\mu^2L\omega_{cr}}\nn 
+ log\frac{\omega_{cr}}{2\mu^2L}\bigg) \\
&+ \frac{M^2L}{3\pi E}\bigg(\frac{\pi^2}{6} - \frac{\omega_{cr}}{E}log\frac{\omega_{cr}}{2\mu^2L} + log\frac{E}{2\mu^2L}\bigg)\bigg],\nn
\eea
where $\omega_{cr}=$ 2.5 GeV. The formalism is not applicable for large quark mass thus we have used it only for charm quarks.
%%%%%%%%%%%%%%%%%%%%%%%%%%%%%%%%%%%%%%%%%%%%%%%%%%%%%%
\subsection*{3. Abir, Jamil, Mustafa, and Srivastava \cite{Abir_Jamil}}
\be
\frac{dE}{dx} = 24 \alpha_s^3 \rho_{QGP} \frac{1}{\mu_g}\left(1 - \beta_1\right) \bigg(\sqrt{\frac{1}{\left(1-\beta_1\right)}log\left(\frac{1}{\beta_1}\right)} - 1 \bigg)\mathcal{F}\left(\delta \right)\nn, 
\ee
where,
\be
%\align
%\begin{align*}
\mathcal{F}\left(\delta \right) = 2\delta - \frac{1}{2}log\bigg(\frac{1 + \frac{M^2}{s}e^{2\delta}}{1 + \frac{M^2}{s}e^{-2\delta}} \bigg) - \bigg(\frac{\frac{M^2}{s}sinh\left(2\delta \right)}{1 + 2\frac{M^2}{s}cosh\left(2\delta \right) + \frac{M^4}{s^2}} \bigg),\nn
%\end{align*}
\ee

\be
\delta = \frac{1}{2}log\bigg[\frac{1}{\left(1-\beta_1\right)}log\left(\frac{1}{\beta_1}\right)\left(1 + \sqrt{1-\frac{(1-\beta_1)}{log\left(\frac{1}{\beta_1}\right)}}\right)^2\bigg],\nn
\ee

\be
s = 2E^2 + 2E\sqrt{E^2 - M^2} - M^2, \  \beta_1 = \mu_g^2 / (C E T),\nn
\ee

\be
C = \frac{3}{2} - \frac{M^2}{4ET} + \frac{M^4}{48E^2T^2\beta_0}log\bigg[\frac{M^2 + 6ET(1+\beta_0)}{M^2 + 6ET(1-\beta_0)}\bigg],\nn
\ee

\be
\beta_0 = \sqrt{1 - \frac{M^2}{E^2}}, \ \rho_{QGP} = \rho_q + \frac{9}{4}\rho_{g},\nn
\ee

\be
\rho_q = 16T^3\frac{1.202}{\pi^2}, \ \rho_g = 9N_fT^3\frac{1.202}{\pi^2}.\nn
\ee
The kinematic factor $\mathcal{F}\left(\delta \right)$ in \textit{Corrected} AJMS formalism is adopted from~\cite{Kapil_NPA}.

In all formalisms, $\mu = \sqrt{4\pi\alpha_sT^2\left(1 + N_f/6\right)}$ is the Debye screening mass, $m_g = \frac{\mu}{\sqrt{2}}$ is 
the thermal gluon mass, $T$ is the temperature of the QGP medium. $\alpha_s$ is the coupling
constant for strong interaction and the value is taken as $0.3$, independent of temperature. $N_f\left(=3\right)$ is the number of active quark flavours.

\section*{Data Availability Statement}
The authors declare that the data supporting the findings of this study are available in public domain and the sources are duly cited accordingly in the manuscript. 




%\end{appendices}




%\section{References}
%\clearpage

\end{document}